\documentclass[useAMS,usenatbib]{mnras}
 \usepackage{amsmath}
  \usepackage{graphicx}
  \usepackage{subfigure}
  \usepackage{comment}
  \usepackage{bm}
  \usepackage{amssymb}
  \usepackage{scalerel}
  \usepackage{esvect}
  \usepackage{xcolor}
  \usepackage{color}
\usepackage{footnotebackref}  
\usepackage{threeparttable}

\newcommand{\kepler}{{\em Kepler}}
\newcommand{\corot}{{\em CoRoT}}
\newcommand{\numax}{\mbox{$\nu_{\rm max}$}}
\newcommand{\Dnu}{\mbox{$\Delta \nu$}}

\newcommand{\muHz}{\mbox{$\mu$Hz}}

\newcommand{\Dpi}{\Delta \Pi}

\newcommand{\gr}{\color{green}\bf}
\newcommand{\msun}{\!{\rm M_{\sun}}}
\newcommand{\rsun}{\!{\rm R_{\sun}}}
\newcommand{\np}{{n_{\rm p}}}
\def\jnote #1]{{\bf\gr #1]}}

\title[Variations of mixed modes in red giant stars]
{Variations of the mixing character of dipolar mixed modes in red giant stars}
\author[C.~Jiang, M. Cunha and J.~Christensen-Dalsgaard]
{\parbox{\textwidth}{C.~Jiang,$^{1,2}$\thanks{E-mail:jiangch53@mail.sysu.edu.cn}
M.~Cunha$^{2,3}$, J.~Christensen-Dalsgaard$^{4}$, QS.~Zhang$^{5,6,7,8}$}\vspace{0.4cm}\\
\parbox{\textwidth}{
$^{1}$School of Physics and Astronomy, Sun Yat-Sen University, No. 135, Xingang Xi Road, Guangzhou, 510275, P. R. China \\
$^{2}$Instituto de Astrof\'{\i}sica e Ci{\^e}ncias do Espa\c{c}o, Universidade do Porto, CAUP, Rua das Estrelas, PT4150-762 Porto, Portugal\\
$^{3}$School of Physics and Astronomy, University of Birmingham,Birmingham, B15 2TT, United Kingdom\\
$^{4}$Stellar Astrophysics Centre, Department of Physics and Astronomy, Aarhus University, Ny Munkegade 120, DK-8000 Aarhus C, Denmark\\
$^{5}${Yunnan Observatories, Chinese Academy of Sciences, 396 Yangfangwang, Guandu District, Kunming 650216, China}\\
$^{6}${Center for Astronomical Mega-Science, Chinese Academy of Sciences, 20A Datun Road, Chaoyang District, Beijing 100012, China}\\
$^{7}$Key Laboratory for the Structure and Evolution of Celestial Objects, Chinese Academy of Sciences, 396 Yangfangwang, Guandu District, Kunming 650216, China\\
$^{8}${University of Chinese Academy of Sciences, Beijing 100049, China}}}

\pagerange{\pageref{firstpage}--\pageref{lastpage}} \pubyear{2020}

\def\LaTeX{L\kern-.36em\raise.3ex\hbox{a}\kern-.15em
    T\kern-.1667em\lower.7ex\hbox{E}\kern-.125emX}

\begin{document}

\label{firstpage}

\maketitle

\begin{abstract}
Thanks to the high quality data of space missions, the detection of mixed modes has become possible in numerous stars. In this work, we investigate how the mixing character of dipolar mixed modes changes with stellar evolution, as well as with frequency within each stellar model. This is achieved by monitoring the variations in the coupling strength and the period spacing of dipolar mixed modes in red-giant models. These parameters are measured by fitting the asymptotic expansion of mixed modes to the model frequencies of a grid of red-giant models with masses between 1.0 and 2.0 $\msun$ and three different chemical abundances. 
The coupling strength and the period spacing decrease with stellar
evolution. We find that the slopes of their decreasing trends depend on the radial order of the pressure mode component. 
A non-negligible increase of the coupling strength with frequency by up to around 40\% is found in the observable frequency range for a set of red-giant models. On the contrary, no significant changes of the period spacing with frequency are found. The changes in the mixing character of the modes are in most cases affected by the model mass and metallicity. Buoyancy glitches also have an impact on the mixing character. Significant fluctuations in the estimated coupling strength and period spacing are found for models approaching the luminosity bump, if the glitch impact of the frequencies is not considered in the applied asymptotic expansion.
\end{abstract}

\begin{keywords}
asteroseismology - stars: interiors - stars: oscillations. 
\end{keywords}

\section{Introduction}

Asteroseismology provides us with a unique tool to study the interior of stars with mixed modes which have a mixed pressure-gravity nature \citep{tak16a, tak16b, mos17}. Thanks to the very long-term and high-precision photometric spaceborne observation missions, such as \corot\ and \kepler, extensive detection of mixed modes in subgiant and red-giant stars  has become possible, which has led to the bloom of the studies of mixed modes \citep[e.g.,][]{bed10, bec11, jcd12}. Mixed modes are of great importance because by determining their gravity-mode (g-mode) period spacing we are able to probe the core \citep{bed11}, and to monitor its rotation \citep{bec12,mos12a} for the first time. The g-mode character carries physical information about the core region along with the oscillation modes, while the pressure-mode (p-mode) character makes them penetrate to the surface area of stars and hence become more detectable.   

Mixed modes are detected in evolved stars past the main sequence, namely in subgiant and red-giant stars. In these stars the very condensed core increases the gravitational acceleration and hence the buoyancy frequency, which opens up the possibility of mixed g- and p-mode character of the oscillations. Such mixed modes in subgiant stars were first observed and modelled in $\eta$ Boo \citep{hans95a, hans03, jcd95, car05} and $\beta$ Hyi \citep{bed07, bra11}. Unlike mixed modes in subgiants where the p-mode character is still dominant, most mixed modes in red giants show a strong g-mode character. Many analyses of oscillations of red-giant stars have been done with \corot\ and \kepler\ data \citep[e.g.,][]{hek09,bed10,hub10,jiang11,mat11,mos11,bau12,kal12}. 

For a solar-like oscillator the large density gradient outside the core divides the star into two cavities: gravity (g-mode) cavity and acoustic (p-mode) cavity. Between the two cavities there is the so-called \textit{evanescent region} where the mode has an exponential behaviour and thus is evanescent. The location, shape and size of the evanescent region may all play a role in the coupling between these two cavities and, thus, in the characteristics of the mixed modes. When coupling between the cavities occurs, the mode frequencies shift from what would be the pure p-mode value, in a series of avoided crossings \citep{osa75,aiz77}. The strength of the coupling between the g- and p-mode cavities is measured by a dimensionless coefficient $q$, the so-called \textit{coupling strength}, which is a key parameter to characterize the mixed modes. 

The characteristics of mixed modes, such as the g-mode period spacing and the coupling strength, have proven to be useful seismic diagnostics of the inner structure of red giants \cite[e.g.,][]{mos12b, mon13, mos15, hek17}. 
The two parameters can be determined by applying pulsation theory to the g-mode character of the mixed modes. \cite{buy16} determined the g-mode period spacings for three red giants observed by \kepler, by means of empirically fitting the observed period spacings and by means of the asymptotic relation. They found compatible results between the two approaches.  
\cite{ben12} fitted mixed modes in sub-giant models and found that the coupling strength of the dipolar mixed modes is predominantly a function of stellar mass and appears to be independent of metallicity. 
\cite{pin20} investigated the potential of the coupling strength in probing the mid-layer structure of red giants, by applying a detailed analytical approach to stellar structure and oscillations. They noted that the value of the coupling strength depends on the radial extent of the evanescent region and the density scale height in that region, which shed some light on understanding the variations of the coupling strength with stellar evolution.
In this paper, we present our analysis of the mixed modes in RGB models by fitting the theoretical mode frequencies to obtain the coupling strength and the g-mode period spacing. We investigate the relation between the variation of the mixing character and stellar properties.

\section{Mixed Modes in Red Giants}

The properties of mixed modes can conveniently be analysed in terms of
the coupling between fictitious pure p and g modes.
In a subgiant star the density of p modes is much higher than the density of g modes. Here, an avoided crossing can be characterized by an $n$-mode system of $(n -1)$ p modes coupled with a low radial-order ($n_{\rm g}$) g mode \citep{deh10}. Thus a mixed mode can be characterized by the orders $\np$ of its p-mode component and $n_{\rm g}$ of its g-mode component, and the resulting order $n_{\rm m}$ of the mixed mode is defined as $n_{\rm m} = n_{\rm g} + n_{\rm p}$ (noting that $n_{\rm g}$ is normally given a negative value). The avoided crossings in subgiants are well displayed in a replicated frequency \'{e}chelle diagram that plots the mode frequencies against themselves modulo the mean large frequency separation $\Dnu$ \citep{gre83, bed12}. During each avoided crossing the oscillation modes gain g-mode character and become g-mode-like mixed modes.

On the other hand, for red giants, the frequencies of g modes increase so that the  p modes are coupled with a dense set of high-$n_{\rm g}$ g modes. In such cases, the mixture of p and g modes makes the original pure g modes gain p-mode character. This can be best illustrated by a period \'{e}chelle diagram \citep{bed11, mos12b, jiang14} in which g-mode period spacing is used instead of $\Dnu$ in the normal frequency \'{e}chelle diagram. This can be done because the periods of g modes with the same mode degree $\ell$ are nearly equally spaced by the period spacing, So the g modes would stack vertically in such an \'{e}chelle diagram. \cite{mos12b} introduced an asymptotic relation for mixed modes, based on the analysis of \cite{shiba79} and \cite{unn89}, as
\begin{equation}
 \nu = \nu_{\np,\,\ell} + \frac{\Dnu}{\pi}\arctan \left[q\tan \pi \left(\frac{1}{\nu \Dpi_{\ell}}-\epsilon \right) \right],
 \label{eq:mixnu}
 \end{equation}
where $\nu_{\np,\,\ell}$ is the uncoupled solution for the p mode with radial order $\np$ and the gravity offset $\epsilon$ is  linked to the properties of the evanescent region in RGB stars. This relation has proven to be extremely powerful in the analysis of mixed-mode properties \citep[e.g.][]{hek18, mos18, pin19}. $\Dpi_{\ell}$ is the g-mode period spacing, for which the asymptotic expression is given by
\begin{equation}
\Dpi_{\ell,\rm{asy}} = \frac{2\pi^2}{\sqrt{\ell(\ell+1)}} \left(\int_{r_1}^{r_2} N_{\rm BV} \frac{\mathrm dr}{r} \right)^{-1},
\label{eq:dpi}
\end{equation}
where $N_{\rm BV}$ is the Brunt-V\"ais\"al\"a frequency (also called the buoyancy frequency) and the integral is over the g-mode cavity. For each $\np$-order p mode, we can assign a group of $\mathcal{N} + 1$ mixed modes where $\mathcal{N} \approx \Delta \nu \Delta \Pi_{\ell}^{-1} \nu^{-2}_{\np,\,\ell}$, with the most p-mode-like modes positioned at the centre. 
The asymptotic expansion in equation~\eqref{eq:mixnu} and its alternative mathematical descriptions have successfully been applied in many studies of the observed mixed modes \citep{mos12b, mos14, buy16} and of the theoretical ones \citep{jiang14, hek17, jiang18, hek18, pin19, pin20}.
However, equation~\eqref{eq:mixnu} does not include the effect of structural glitches in the core that is discussed extensively by \cite{cunha15} and \cite{cunha19}. 
In the following analysis, we employed a method based on Bayesian nested sampling to obtain the coupling strength and the period spacing. Since we focus on only dipolar ($\ell =1$) mixed modes, hereafter we will simply use $\nu_{\np}$ and $\Dpi_1$ to represent $\nu_{\np,\,\ell = 1}$ and $\Dpi_{\ell = 1}$. We first discuss the variations of $q$ and $\Dpi_1$ with stellar properties, and then further explore the effect of glitches on these variations.

\section{Models}
\label{sc:models}
We have computed a series of stellar models to investigate the properties of mixed modes. The theoretical model grid was generated using the {\scriptsize ASTEC} evolution code \citep{jcd08a} and the corresponding oscillation frequencies with the {\scriptsize ADIPLS} oscillation package \citep{jcd08b}. Models with masses in the range of 1.0--2.0 $\msun$ were generated with a step of 0.2 $\msun$. All models were computed for three different chemical abundances: the metallicity calibrated to the sun ($Z = 0.0173, [{\rm Fe/H}] = 0$), a higher metallicity ($Z = 0.0295, [{\rm Fe/H}] = 0.25$) and a metal-poor case ($Z=0.0099, [{\rm Fe/H}]=-0.25$). Convection was treated under the assumption of mixing-length theory \citep{boh58} with the mixing-length parameter set to 1.96 that is calibrated to solar models. The input physics of the current modelling included the latest OPAL opacity tables \citep{igl96}, OPAL equation of state in its 2005 version \citep{rog96}, and NACRE reaction rates \citep{ang99}. The models were kept as simple as possible, neglecting overshooting, diffusion and rotation.

\begin{figure*}
\resizebox{0.8\hsize}{!}{\includegraphics[angle=0]{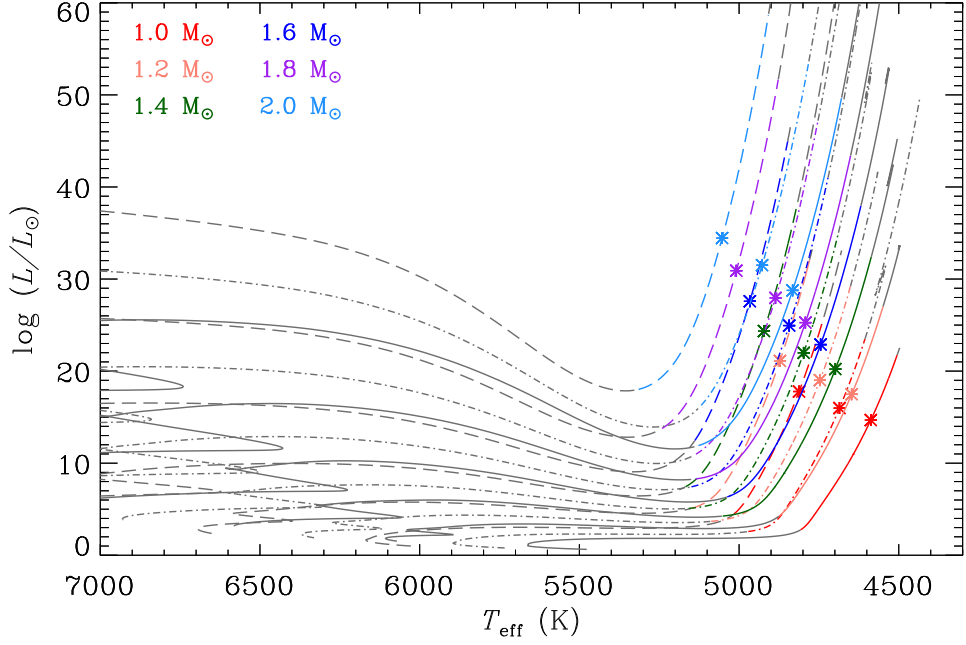}}
\caption{Evolutionary tracks (dark grey) for all model sequences considered in this work, with three different metallicities, the metal-poor case (dashed), the solar value (dash dot) and the metal-rich one (solid), and with six masses indicated by the colours on their RGB. The coloured sections indicate the region where models are selected for the analysis of Sections~\ref{sc:cpvar} and~\ref{sc:psvar}. The asterisk symbols, coloured according to the colour coding of the tracks, indicate the locations of models with $\Dnu \approx 9 \, \muHz$, which are discussed in Section~\ref{sc:cpvarfre}. }
\label{fg:tracks}
\end{figure*}

In total we computed 18 evolutionary tracks, shown in Figure~\ref{fg:tracks}, starting from the zero-age main sequence to the later stage of the red-giant branch (RGB) with hydrogen burning in the shell outside the helium core. Along the evolutionary tracks, we mainly focus on the mixed modes on the RGB. The selection of models for frequency calculations starts from the transit point between the subgiant branch (SGB) and the RGB to a later stage on the RGB where the models have $\numax \approx 60 \,\muHz$ with $\numax = (M / \msun) (R / \rsun)^{-2} (T_{\rm eff} / 5777\,{\rm K})^{-0.5} \, 3050 \, \muHz$ \citep{hans95b} being the estimated frequency of maximum oscillation amplitude. Adiabatic oscillation modes were calculated with the full set of the differential equations, without assuming the Cowling approximation \citep{cow41}. 

\section{Fitting Method}
\label{sc:fittingMethod}
We concentrate on mixed modes in RGB stars where the g-mode behaviour is dominant. When a g mode gains p-mode character, the mode frequency also changes. Therefore, the period spacing $\Dpi$ between two consecutive $n_{\rm g}$ mixed modes differs from $\Dpi_1$. 
In a period \'{e}chelle diagram, mixed modes gaining p-mode character would shift sideways, breaking the vertical alignment of the g modes, with the most p-mode-like one moving farthest.
On an observed power spectrum the most p-mode-like one usually has the highest amplitude in the same mixed mode peak forest (modes with same $\np$ and experiencing the same avoided crossing) and its frequency is the closest to $\nu_{\np}$. Mixed modes in the same peak forest satisfy equation~\eqref{eq:mixnu} with the same $q$ and $\Dpi_1$. Therefore, we may estimate the parameters for each avoided crossing to monitor their relations with $\nu_{\np}$.

The estimation is carried out by fitting the mode frequencies to equation~\eqref{eq:mixnu}. In order to make this process robust and consistent, we apply the Bayesian-nested-sampling tool {\scriptsize DIAMONDS} \citep[high-DImensional And multi-MOdal NesteD Sampling,][]{cor14} which performs Bayesian parameter estimation and model comparison by means of the nested sampling Monte Carlo \citep[NSMC,][]{sk04} algorithm. According to Bayes' theorem the posterior probability is:
\begin{equation}
P(\bm{H} | E) = \frac{\mathcal{L}(E | \bm{H}) \cdot P(\bm{H})}{P(E)} 
\label{eq:bayes}
\end{equation}
where $\bm{H}$ stands for the {\it hypothesis} that is, in this analysis, the free parameter vector $\bm{H} = (q, \Dpi_1, \epsilon, \nu_{\np}$), and {\it evidence} $E$ corresponds to the model frequencies $\nu_{\rm mod}$. $P(\bm{H})$ is the prior that expresses our knowledge about the hypothetical parameters. The denominator $P(E)$ is a normalization factor, generally termed as marginal likelihood or Bayesian evidence. 
$\mathcal{L}(E | \bm{H})$ is the likelihood function that needs to be maximized in the sampling process and is defined as
\begin{equation}
\ln [\mathcal{L}(E | \bm{H})] = - \frac{1}{2} \sum_n \frac{(\nu_{\rm mod} (n) - \nu(n))^2}  {\sigma^2_n},
\label{eq:loglikeli}
\end{equation}
with $\nu$ being the mixed mode frequencies calculated using the asymptotic equation~\eqref{eq:mixnu}. and $\sigma^2_n$ being the frequency errors that are set to be 1 for our model frequencies. By comparing $\nu_{\rm mod}$ and $\nu$ the likelihood is maximized in each iteration.
In equation~\eqref{eq:loglikeli} the likelihood is expressed in logarithm and accordingly the logarithmic posterior becomes
\begin{equation} 
\ln [P(\bm{H} | E)] = \ln [\mathcal{L}(E | \bm{H} )] + \ln[P(\bm{H})] - \ln[P(E)].
\label{eq:logpost}
\end{equation}
Having the likelihood defined and the model data provided, the Bayesian tool also needs the expression for the prior $P(\bm{H})$. We choose uniform prior which requires hyper parameters defining the lower and upper bounds for the free parameters. The hyper parameters define the search space of the prior distribution. The coupling strength is theoretically between 0 and 1 so the hyper parameters for it are set to be 0.01 and 0.99. For $\Dpi_1$, $\epsilon$ and $\nu_{\np}$, solutions are searched within the range of [$\Dpi_{\rm g} - 1.0$ s, $\Dpi_{\rm g} + 1.0$ s], [-0.5, 0.5] and [$\nu_{\rm p-m} - 1.0~\muHz$, $\nu_{\rm p-m} + 1.0~\muHz$], respectively. $\Dpi_{\rm g}$ is the initial guess of $\Dpi_1$ and is obtained by aligning the mixed modes vertically in the period \'{e}chelle plot \citep{bed11}. Also, $\nu_{\rm p-m}$ is the preliminary guess of $\nu_{\np}$, the value of which is achieved by locating the minima of the mode inertia of the mixed modes \citep{jiang14}.  The determinations of $\Dpi_{\rm g}$ and $\nu_{\rm p-m}$ give very good estimates of  $\Dpi_1$ and $\nu_{\np}$ when the coupling is not extremely weak. After many tests, we found that the adopted searching space of the parameters is large enough to return stable outputs. 

\begin{figure}
\resizebox{1.0\hsize}{!}{\includegraphics[angle=90]{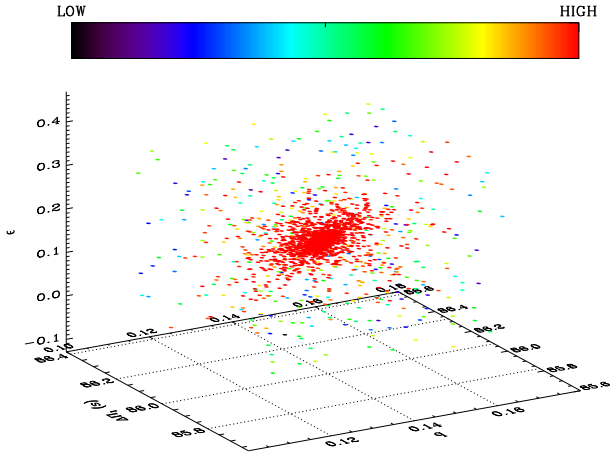}}
\caption{Sampling points drawn in the three-dimensional hyperp-parameter space by the NSMC process, for the 1.0 $\msun$, $Z = 0.0295$ model with $\numax = 177.81\, \muHz$ model, colour-coded according to the logarithm of the likelihood values.}
\label{fg:sampling_points}
\end{figure}

\begin{figure*}
\resizebox{1.0\hsize}{!}{\includegraphics[angle=0]{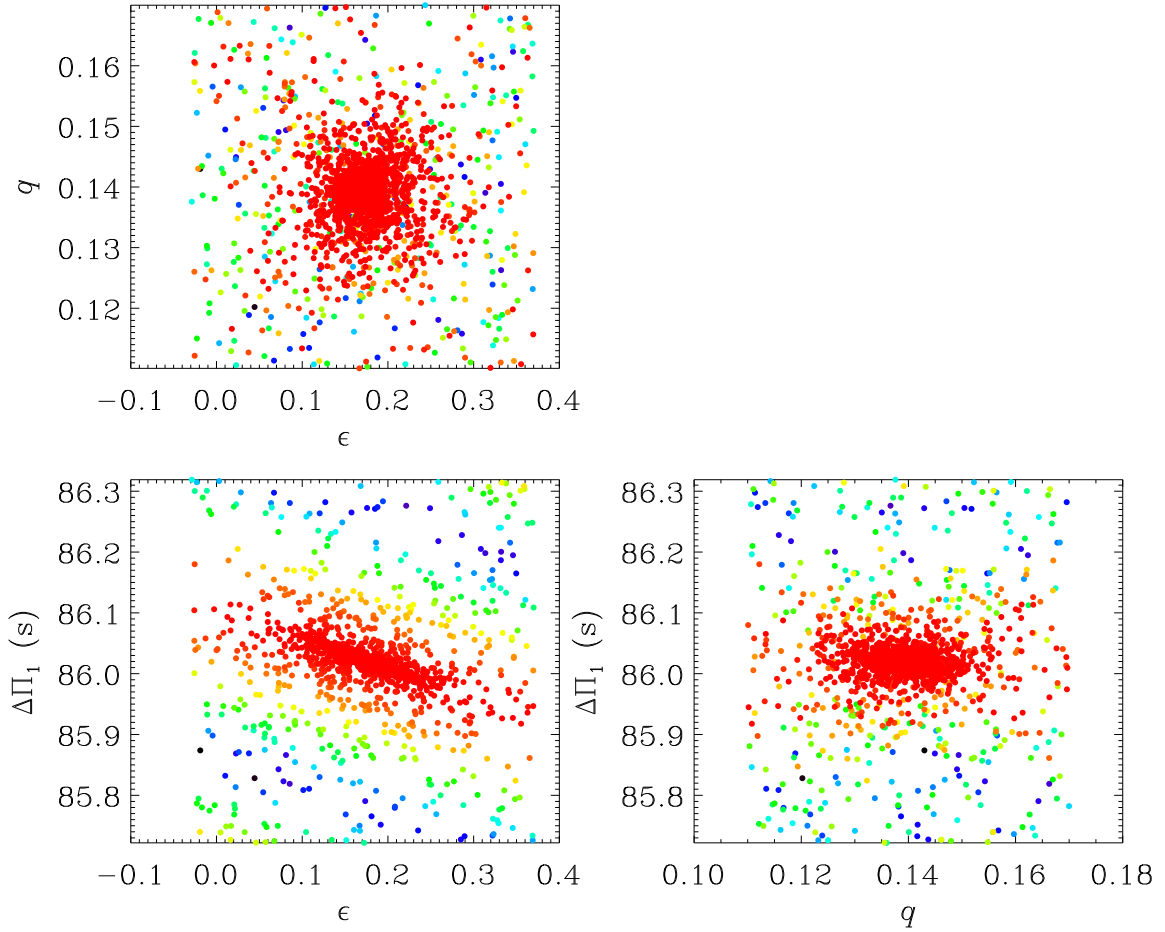}}
	\caption{Correlation maps indicating the logarithm of the likelihood values for the same model as shown in Figure~\ref{fg:sampling_points}. The same colour code of Figure~\ref{fg:sampling_points} is used for the symbol colours. \textit{Top left}: correlation map of the two-dimensional parameter space ($q, \, \epsilon$), the reddest points indicate the most likely solution. \textit{Bottom left}: correlation map of the two-dimensional parameter space ($\Dpi_1, \, \epsilon$). A strong correlation is observed between these two parameters. \textit{Bottom right}: correlation map of the two-dimensional parameter space ($\Dpi_1,\,q$). }
\label{fg:mapping}
\end{figure*}

The detailed sampling and computations of {\scriptsize DIAMONDS} are introduced in \cite{cor14}. An example of the sampling points drawn by the NSMC process is shown in Figure~\ref{fg:sampling_points} for the 1.0 $\msun$, $Z = 0.0295$ model with $\numax = 177.81\, \muHz$ that has a biggest $\ln [\mathcal{L}(E | \bm{H})]$] value of -0.005. For the illustrated model, around 2500 points are drawn in the parameter space before the nested sampling hits the stopping criterion. The points with high likelihood are centred in the parameter space and give the best fits to the theoretical frequencies. Figure~\ref{fg:mapping} illustrates the different distributions of $\ln [\mathcal{L}(E | \bm{H})]$ for the parameters estimated by equation~\eqref{eq:mixnu} and our fitting procedure, for the model shown in Figure~\ref{fg:sampling_points}. The correlation maps show that there is a significant correlation between $\Dpi_1$ and $\epsilon$ (bottom left panel of Figure~\ref{fg:mapping}), which is consistent with the findings of \cite{buy16}. They also fitted the mixed modes with equation~\eqref{eq:mixnu} by means of a grid-search method to derive the parameters ($\Dpi_1,\, q, \, \epsilon$), and discussed in detail the correlation between $\Dpi_1$ and $\epsilon$.

\begin{figure}
\resizebox{1.0\hsize}{!}{\includegraphics[angle =90]{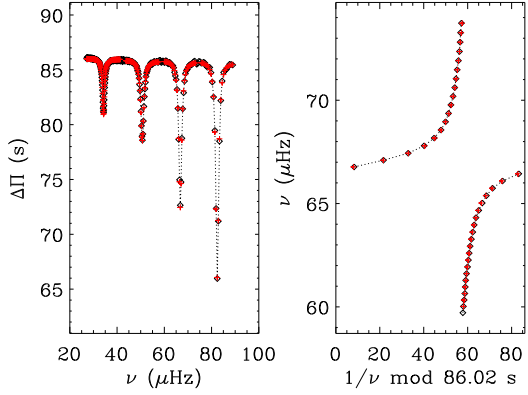}}
	\caption{Fitting results for the same model as shown in Figure~\ref{fg:sampling_points}. \textit{Left}: Period spacings $\Dpi$ between two consecutive $n_{\rm g}$ mixed modes against the mode frequencies. Model values are coloured in black, while the best-fit values are in red. \textit{Right}: Period \'{e}chelle diagram for the third avoided crossing shown in the left figure, using the same notation. The values estimated for the parameters are $q = 0.139$ and $\Dpi_1= 86.02 \, \rm s$. }
\label{fg:best-fit}
\end{figure}

The fitting is separately performed for each avoided crossing centred on $\nu_{\np}$, using modes with frequencies in the range [$\nu_{\rm p-m} - \Dnu / 2, \nu_{\rm p-m} + \Dnu / 2$] for each run.  
The effectiveness of the fit is mainly determined by the mixed-mode density $\mathcal{N}$. A small number of modes results in unstable estimations for the parameters. We select avoided crossings with a number of modes larger than 5 in the fitting process, which rules out very high-$\np$ modes.
$\Dnu$ is determined as a mean value, obtained by performing a linear fit to 7 radial modes around $\numax$. However, $\Dnu$ is not strictly constant across the power spectrum. It becomes slightly smaller for low-$\np$ modes. In this case, modes at the edge of the frequency range [$\nu_{\rm p-m} - \Dnu / 2, \nu_{\rm p-m} + \Dnu / 2$] would couple with adjacent $\np$ modes. 
Therefore, a smaller frequency range of 0.5 $\Dnu$, with a width of 0.25 $\Dnu$ on each side of $\nu_{\rm p-m}$, are used for low-$\np$ modes in the fitting process. 
The reduction of the frequency range is determined by manually examining the vertical pattern of the g-mode-like mixed modes in the period \'{e}chelle diagram. These modes would not stack vertically in the diagram if the fitting range should be reduced.
We find only very low-$\np$ modes in our evolved models need a reduced frequency range. Since the density of these low-$\np$ modes is very high, a smaller frequency range does not affect the estimations of $q$ and $\Dpi_1$.
In general, our fitting procedure yields good estimations of $q$ and $\Dpi_1$. Figure~\ref{fg:best-fit} shows the fitting results for the model presented in Figures~\ref{fg:sampling_points} and~\ref{fg:mapping}, which demonstrates excellent matching between the model frequencies and frequencies calculated using equation~\eqref{eq:mixnu} with the fitted parameters. The range of $\ln [\mathcal{L}(E | \bm{H})]$ of the best fits for all the models is between -1.84 and -0.0001, with an average value of -0.11.

Young models close to the SGB have strong coupling \citep{tak16b, mos17}, which makes their low-$\np$ modes hard to fit, for two reasons. Firstly, the local minima of the mode inertia is less pronounced in these stars, making it difficult to locate the frequencies $\nu_{\rm p-m}$ which are necessary to fit equation~\eqref{eq:mixnu} to the model data. Secondly, the mixed-mode density $\mathcal{N}$ of these less evolved models is small for modes with frequencies around $\numax$, so the parameters inferred from fitting equation~\eqref{eq:mixnu} to data have large uncertainties \citep{mos17}. Therefore, SGB models are not used in the analysis.

With the fitting method introduced in this section, we may trace the variations of $q$ and $\Dpi_1$ as the model evolves and compare them for models with different properties. 

\section{Coupling Strength Variations}
\label{sc:cpvar}

As the star evolves on the RGB, the size of the evanescent region inevitably changes as the inner structure of the star changes. Thus, the value of $q$ is expected to vary with stellar evolution \citep{mos17, hek18}. Furthermore, $q$ is also frequency dependent because the size of the evanescent region varies with mode frequency. The frequency dependence of $q$ was first expressed in the pioneering work of \cite{shiba79} and has been discussed in many recent publications, such as \cite{jiang14, hek18, cunha19}. However, in fits to observed frequencies $q$ is normally assumed to be constant across the observable frequency domain of solar-like oscillations; this is mainly due to the difficulty to extract the frequency dependence of $q$ given the quality of the observational data.
In this section, using our models we explore the dependence of $q$ on stellar evolution and mode frequency.

\subsection{Variation of $q$ with stellar evolution}
\label{sc:cpvarage}

\begin{figure}
\resizebox{1.0\hsize}{!}{\includegraphics[angle =90]{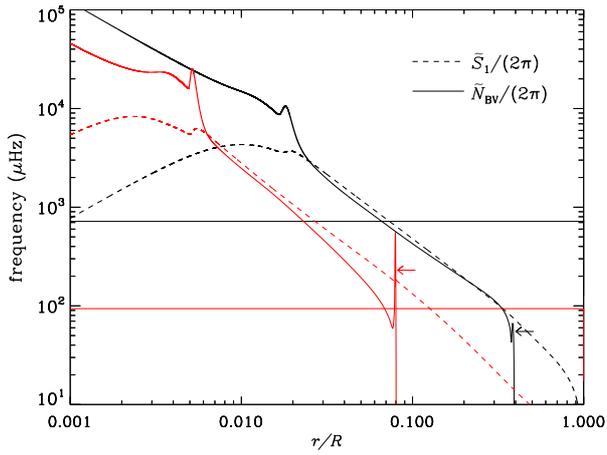}}
	\caption{Propagation diagram showing $\tilde{N}_{\rm BV}/2 \pi$ and $\tilde{S}_{\rm 1}/2 \pi$ for two 1.0 $\msun$ models with solar metallicity, one early RGB (black, $\Dnu = 43.16~\muHz$), and one more evolved RGB (red, $\Dnu = 8.96~\muHz$) model that is marked by the red asterisk symbol on the dash-dotted red curve in Figure~\ref{fg:tracks}. The $\tilde{N}_{\rm BV}$ are indicated by solid curves and the $\tilde{S}_1$ by dashed curves. The values of $\numax$ for each model are indicated by the horizontal lines. The spikes (marked by the arrows) in the buoyancy frequency of both models are caused by the hydrogen discontinuity discussed in Section~\ref{sc:glitch}.} 
\label{fg:propagation}
\end{figure}

The properties of mixed modes depend on characteristic gravity-wave and
acoustic frequencies, conveniently illustrated in a propagation diagram.
Figure~\ref{fg:propagation} illustrates the resulting propagation diagram
for $\ell = 1$ modes.  The cases of an early RGB model located right after the transition region between the SGB and the RGB, and a more evolved RGB model are shown.
\cite{tak06} showed that adiabatic dipolar oscillations of stars can be described by a second-order system of ordinary differential equations with no singularity (except at the boundaries), which is derived without the assumption of the Cowling approximation. In the second-order system the properties of the oscillations are characterized by modified versions of the buoyancy frequency $\tilde{N}_{\rm BV}$ and the Lamb frequency $\tilde{S}_{\rm 1}$. Since our models are computed without assuming the Cowling approximation, $\tilde{N}_{\rm BV}$ and $\tilde{S}_{\rm 1}$ are used in the analysis and plotted in Figure~\ref{fg:propagation}.
The propagation region of g modes is determined by $\tilde{N}_{\rm BV}$ and $\tilde{S}_{\rm 1}$.
In particular, the inner turning point of low-frequency g-modes are constrained by $\tilde{S}_{\rm 1}$, whereas it is set by the original buoyancy frequency in the Cowling approximation. The horizontal lines in Figure~\ref{fg:propagation} indicate the values of $\numax$ around which solar-like oscillations are likely to be detected. The early RGB model shows parallel variations of $\tilde{N}_{\rm BV}$ and $\tilde{S}_{\rm 1}$ with radius in a frequency range around $\numax$. As the model evolves on the RGB, the stellar core contracts and the envelope expands so that the size of the evanescent region expands with $\tilde{N}_{\rm BV}$ and $\tilde{S}_{\rm 1}$ being no longer parallel. See also \citet{pin19, pin20} for detailed analyses of the properties of the oscillations in terms of this formulation.

Due to the expansion of the evanescent region we can expect a decrease in $q$ as the star evolves. The decrease of $q$ is indeed found by \cite{mos17} and \cite{hek18} based on the analysis of several RGB models, despite the different asymptotic expansions and approaches they used to measure the coupling strength. Following these previous studies, we make use of our large sample of models to investigate the variation of $q$ with stellar evolution. To obtain a better theoretical understanding of the dependence of the oscillation properties on stellar evolution, we include in the analysis modes of low acoustic order, even though they are outside the typical range of observed frequencies.
 
\begin{figure*}
\resizebox{1.0\hsize}{!}{\includegraphics[angle =90]{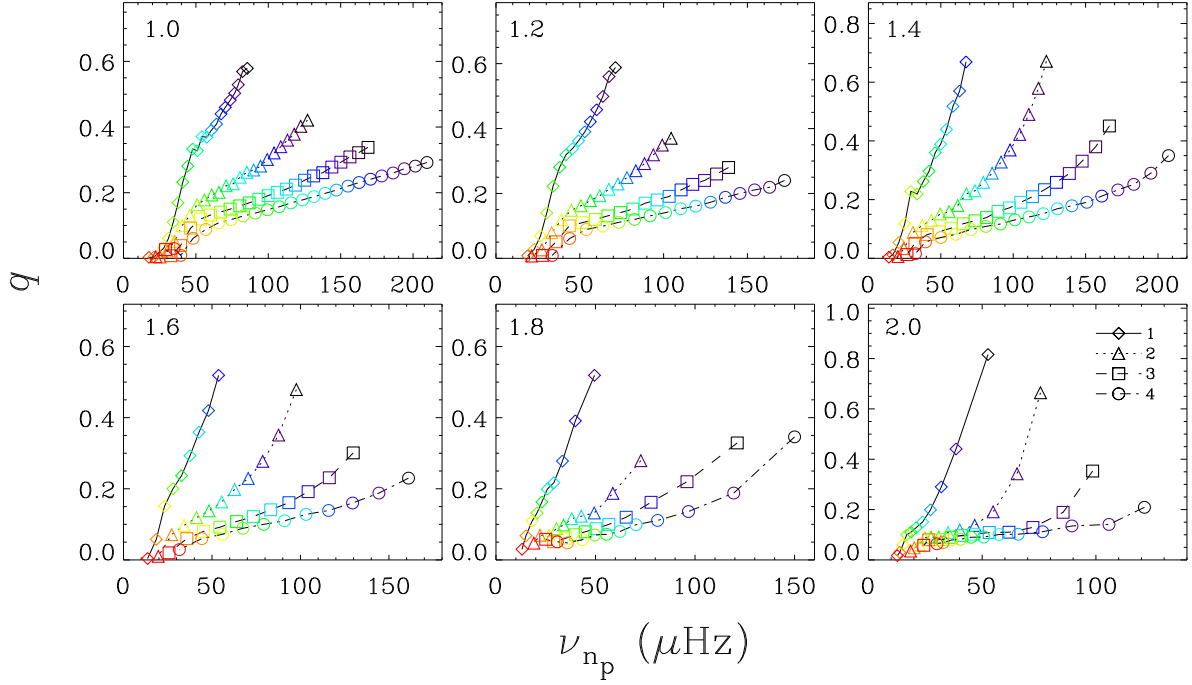}}
	\caption{Variation of $q$ with stellar evolution, for low-$\np$ (1 to 4) modes in models with different masses (values shown in units of solar mass in the upper left corner of each diagram) at $Z = 0.0295$. The acoustic resonance frequency $\nu_{\np}$ of an $\np$ order mode decreases with model age, so the evolution goes from high $\nu_{\np}$ to low $\nu_{\np}$. Model ages are also indicated by the colours, with darker colours being the younger models and warmer colours being the older ones. The models are plotted equally spaced in age (details of the age spacing given in Section~\ref{sc:cpvarage}). Modes of the same $\np$ are indicated by the same symbol and linked by the same line. Notations of symbols and line types for each $\np$ are given in the last diagram.} 
\label{fg:low_qvar295}
\end{figure*}

The analysis is done by monitoring the evolution with age of the first four avoided-crossing occurrences ($\np = 1, 2, 3, 4$), as well as the observable modes that have frequencies around $\numax$ ($\np = 10, 11, 12, 13$). We obtain $\nu_{\np}$ and $q$ for each avoided crossing using the method mentioned in Section~\ref{sc:fittingMethod}. The resulting variations of $q$ as function of the $\nu_{\np}$ are shown in Figures~\ref{fg:low_qvar295} for the low-$\np$ modes in $Z=0.0295$ models. The model age is indicated by the symbol colours so that age increases from right to left in each diagram. 
The modes with the same $\np$ are linked by the same line with equal age spacing. For models with mass $ \leq 1.6 ~ \msun$, the age difference between two connected modes is 0.02$\,$Gyr. For the 1.8 and 2.0 $\msun$ models, age intervals of  0.01 and 0.0025$\,$Gyr are used, respectively, to ensure an adequate number of models in the diagram.
We confirm that $q$ decreases as the models climb on the RGB for these low-$\np$ modes. The decreasing trend is universal in that it is present regardless of the model's mass and metallicity (results from models with the other two metallicities considered are shown below in Figures~\ref{fg:qvarmass} and \ref{fg:qvarz}). However, we note that the rate at which $q$ decreases depends on $\np$ of the mixed modes. In all cases, $q$ of the $\np =1$ mode is the highest at the earliest evolutionary stage and decreases fastest along the evolution. The typical uncertainties on $q$ and $\nu_{\np}$ obtained from the fitting process are 0.005 and $0.05 \,\muHz$, respectively, which are small compared with the range of the variations. Therefore both the decreasing trend and its dependence on $\np$ are unambiguous. 

\begin{figure*}
\resizebox{1.0\hsize}{!}{\includegraphics[angle =90]{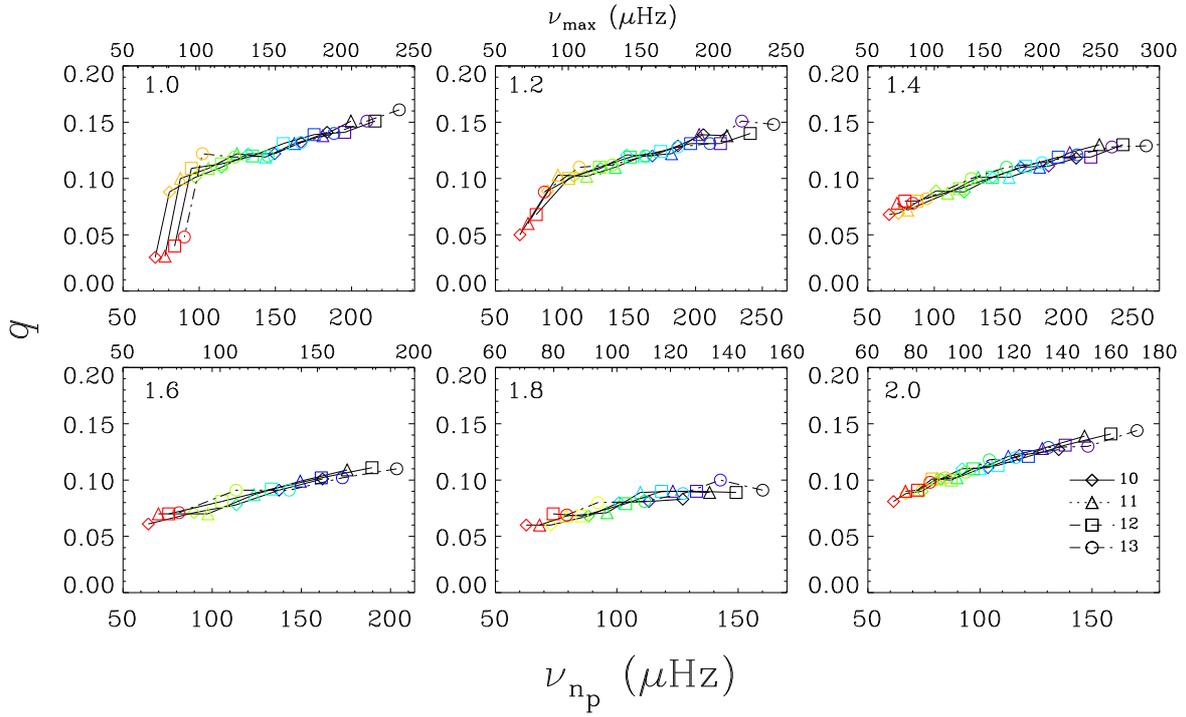}}
\caption{Variation of $q$ with stellar evolution for high-$\np$ (10 to 13) modes in $Z=0.295$ models. Plot settings are the same as in Figure~\ref{fg:low_qvar295}. The axis above each diagram is $\numax$.} 
\label{fg:high_qvar295}
\end{figure*}

As $\np$ increases, the decrease of $q$ becomes more gradual. Therefore, we can expect smaller variation of $q$ with time for high-$\np$ modes, such as those modes in the observable frequency range. Evolution trends of $q$ for modes in the $Z=0.0295$ models with frequencies around $\numax$ are illustrated in Figures~\ref{fg:high_qvar295}, but in a smaller range of $q$ than in Figures~\ref{fg:low_qvar295}. The trends are all entangled indicating similar decreasing rates. Besides, the trends are all relatively gradual, except for the oldest models in the 1.0 and 1.2 $\msun$ series, for which $q$ start to drop sharply. This is on account of a comparable $\numax$ and $\tilde{N}_{\rm BV}$ at the base of the convective zone, similar to what we see on the more evolved RGB model in Figure~\ref{fg:propagation}, which makes the hypothesis of parallel variations of $\tilde{N}_{\rm BV}$ and $\tilde{S}_{\rm 1}$ no longer valid \citep{mon13, mos17, pin19}. In fact, the size of the evanescent region for these evolved models is relatively large and hence the coupling is very weak. If more evolved models are included in the figure, the drop of $q$ will also be seen for higher-mass models.

The decrease of $q$ with stellar evolution is also dependent on the model mass and metallicity. In Figures~\ref{fg:qvarmass} and \ref{fg:qvarz} the decreasing trends for all models with $Z=0.0295$, 0.0173 and 0.0099 are plotted with different arrangements of the model mass and metallicity (details of the arrangements given in the captions of the figures) to reveal their relations. Regardless of metallicity the $q$ of the models with larger masses decreases faster with increasing age than those with smaller masses, as shown in Figure~\ref{fg:qvarmass}. But the difference in the rate of decrease associated with the model masses becomes smaller for higher-$\np$ modes and for more evolved models. As shown in Figure~\ref{fg:qvarz}, the $q$ of less metallic models decreases faster. These mass and metallicity dependencies on $q$ indicate 
that as the star evolves on the RGB the expansion of the evanescent region is also connected with these stellar properties.

\subsection{Variation of $q$ with frequency}
\label{sc:cpvarfre}

In an observational study of mixed modes, the number of observable mixed modes is limited and the accurate measurement of $q$ for each avoided crossing is not possible. Therefore it is common to assume that $q$ is independent of mode frequency so a mean value of $q$ can be used in the analysis. The assumption of a constant $q$ is considered in many works. For instance, \cite{mos17} measured $q$ in about 6100 red giants and a few dozen subgiants, using an automated tool developed by \cite{vra16}, with the assumption of parallel radial variations for $\tilde{N}_{\rm BV}$ and $\tilde{S}_{\rm 1}$ and thus of $q$ being constant in the observable frequency range. However, \cite{mos17} also found that for a limited number of stars a constant $q$ did not provide a satisfying fit of the mixed-mode pattern and for these stars a better fit was obtained when varying $q$ with frequency. The assumption of parallel radial variations for $\tilde{N}_{\rm BV}$ and $\tilde{S}_{\rm 1}$ is based on the asymptotic analysis by \cite{tak16a}, which is inspired by models with $\Dnu > 20~\muHz$, whereas it becomes questionable for more evolved RGB models with $\Dnu < 15~\muHz$ such as those analysed by \cite{hek18} who find different values of $q$ for oscillation modes with different frequencies using the asymptotic formalism of $q$ from \cite{tak16a}. As pointed out before, the propagation diagram in Figure~\ref{fg:propagation} shows similar radial variations of $\tilde{N}_{\rm BV}$ and $\tilde{S}_{\rm 1}$ for the younger model ($\Dnu = 43.16~\muHz$), while for the more evolved model ($\Dnu = 8.96~\muHz$) the two characteristic frequencies approach each other as the radius decreases and the frequency increases, which is consistent with the findings of \cite{hek18}. \cite{mos17} state that for more evolved RGB models, $\numax$ becomes comparable to or even smaller than the value of $\tilde{N}_{\rm BV}$ just below the base of the convective zone \citep[see also][]{pin19}, so that the hypothesis of parallel variations of $\tilde{N}_{\rm BV}$ and $\tilde{S}_{\rm 1}$ is no longer valid. This is also corroborated by the more evolved model shown in Figure~\ref{fg:propagation}. \cite{cunha19} also provided clear evidence for the need of a frequency-varing $q$ from a detailed fitting of a single evolved RGB star. With the help of RGB models, in this section we examine in detail how $q$ varies with the frequency.

\subsubsection{In the observable frequency range}
\label{sc:cpvarhighfre}

From the 18 evolution tracks computed for this analysis, we selected RGB models that have $\Dnu$ around 9.0 $\muHz$ for each track to check the frequency dependence of $q$ for detectable oscillation modes. Since $\Dnu$ is closely related to the stellar density, it becomes smaller as the star climbs up on the RGB when the star expands its radius remarkably and becomes less dense. Models having the same $\Dnu$ but different initial masses are located differently on the evolutionary tracks, meaning that the model structures are also different. Moreover, $\Dnu \approx 9~\muHz$ guarantees that all the models considered here are not young models close to SGB (cf. Figures~\ref{fg:tracks}), so that the mixed modes around $\numax$ are dense enough for the fitting process. In this way, we may examine the frequency dependence of $q$ for high-$\np$ modes from models with different structure. 

\begin{figure}
\resizebox{1.0\hsize}{!}{\includegraphics[angle =90]{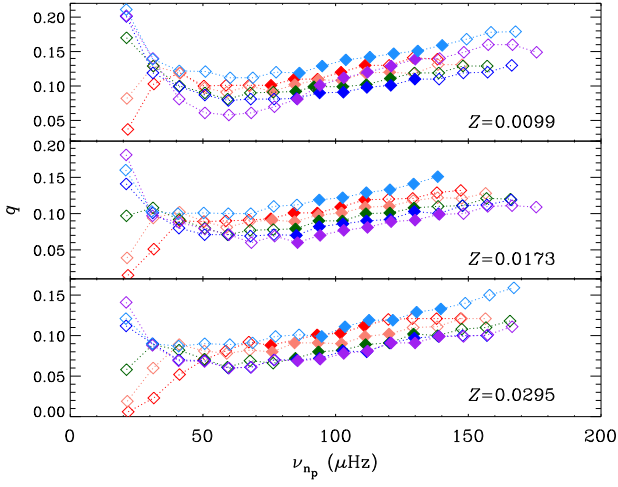}}
	\caption{Variation of $q$ with frequency $\nu_{\np}$, starting from $\np = 1$, for models with three different metallicities and six masses, but with same $\Dnu \approx 9 \, \muHz$. The model mass is indicated by the symbol colour (colour notation given in Figure~\ref{fg:tracks}). Each symbol represents the result estimated from the avoided crossing centred at the given $\nu_{\np}$. The $\numax$ of the models are around 100 $\muHz$. The filled symbols are modes within the detectable solar-like frequency range. }
\label{fg:qdif295}
\end{figure}

The results for the $\Dnu \approx 9.0 \, \muHz$ models are illustrated in Figure~\ref{fg:qdif295} where $q$ is plotted against $\nu_{\np}$ with $\np$ starting from 1. Unlike in the last section, here the modes connected by a line are from the same model but with different $\np$. Therefore, Figure~\ref{fg:qdif295} shows the frequency dependence of $q$. For high-frequency modes $q$ generally increases as the frequency increases, as expected from the fact that the evanescent region becomes narrower for those modes. We select oscillation modes (filled symbols) with frequencies in the expected solar-like domain, defined as $0.75 < \nu / \numax < 1.25$ \citep{mon13}, where $\numax$ is again estimated from the scaling relation in \cite{hans95b}. For these modes connected by each line the pairs $q$ and $\nu_{\np}$ are fitted with a linear formalism adopted by \cite{cunha19},
\begin{equation}
	q = q_1 [\alpha (\nu_{\np} /  \numax - 1) + 1].
\label{eq:qlinfit}
\end{equation}
Equation~\eqref{eq:qlinfit} can be easily transferred to the normal linear model $q = A + B \nu_{\np}$ with,
\begin{align}
A &= -q_1 \alpha + q_1, \label{eq:a} \\
B &= q_1 \alpha / {\numax} . \label{eq:b}
\end{align}
The zero point $A$ is different for different models and is not our concern. We focus on the slope $B$ here.
The inferred parameters $q_1$, $\alpha$, $A$ and $B$ are given in Table~\ref{tb:qvarfit}.
The slopes $B$ are very alike, as anticipated from the visual inspection of Figure~\ref{fg:qdif295}, and confirmed by the values of $B$ given in Table~\ref{tb:qvarfit}, except for the 1.80 $\msun$ metal-poor model that has a significantly larger $B$ and hence is not included in the following analysis. The mean slopes for modes with different masses but the same $Z$ are given in Table~\ref{tb:qvarZ}, while those for modes with different $Z$ but the same masses are given in Table~\ref{tb:qvarM}. These indicate no apparent $Z$ or mass dependence of the frequency increase of $q$. It is worth noting that
the results of 
the RGB-1 model (with $\numax = 105\, \muHz$) analysed by \cite{cunha19} and our 1.0 $\msun$, solar-metallicity model (with $\numax \approx 93 \,\muHz$) are consistent. \cite{cunha19} found $\alpha = 0.692 \pm 0.049$ for the RGB-1 model, which agrees with our result of $\alpha = 0.607 \pm 0.069$ within the errors, though the values of $q_1$ are slight different.
The linear fit yields a mean relation of $q = A + (6.5 \times 10^{-4} \,\muHz^{-1})\,\nu_{\np}$ for these modes around $\numax$ for all $\Dnu \approx 9\, \muHz$ models, which corresponds to an average increase in $q$ of 37\% in the observable frequency range.

From the above analysis, we conclude that for evolved RGB stars, a frequency-dependent $q$ should be considered in the range of observed modes, otherwise deviations in the estimated parameters may result when fitting the mixed modes. Here, the variation of $q$ with frequency was investigated based on models with fixed $\Dnu$. In the future, detailed investigations concerning variation of $q$ in RGB stars within a broader $\Dnu$ range should be conducted, but that is beyond the scope of this paper. 

\subsubsection{In the low-frequency range}
\label{sc:cpvarlowfre}

If models of the same age are examined, we see from Figure~\ref{fg:low_qvar295} that, for the low-frequency modes, significant frequency dependence of $q$ can be found in the young models. The parallel variations of $\tilde{N}_{\rm BV}$ and $\tilde{S}_{\rm 1}$ are not valid for these low-$\np$ modes with frequencies well below $\numax$. For these modes an increase of the width of the evanescent region with frequency exists, which accounts for the frequency dependence seen in the young models in Figure~\ref{fg:low_qvar295}.

Regardless of metallicity, in Figure~\ref{fg:qdif295} we also noticed large values of $q$ and a decrease of $q$ with increasing $\nu_{\np}$ for some low-frequency modes in high-mass models. This is consistent with the results of \cite{mos17} and \cite{hek18}, who found that the coupling strengths of evolved RGB models have larger values. The reason for the large $q$ is still uncertain, but can be related to the fact that 
the lower boundary of the evanescent region coincides with the base of the convective envelope for low-$\np$ modes. 
Our models are not as evolved as theirs, so this only happens to low-frequency modes in our high-mass models.
On the contrary, in low-mass models, the values of $q$ for the low-frequency modes are comparatively small, which is caused by the so-called \textit{buoyancy glitch} that induces variations in the oscillation frequencies, as well as in the inferred $q$, when the star evolves toward the luminosity bump. A detailed analysis of the buoyancy-glitch effects will be presented in Section~\ref{sc:glitch}. 

\section{Period Spacing Variations}
\label{sc:psvar} 

Since dipolar mixed modes are observed abundantly in RGB stars,  it is reasonable to examine also how $\Dpi_1$ varies with stellar evolution, as well as with frequency. With $\Dpi_1$ obtained in our fitting process, we are able to monitor these variations.

\begin{figure*}
\resizebox{1.0\hsize}{!}{\includegraphics[angle =90]{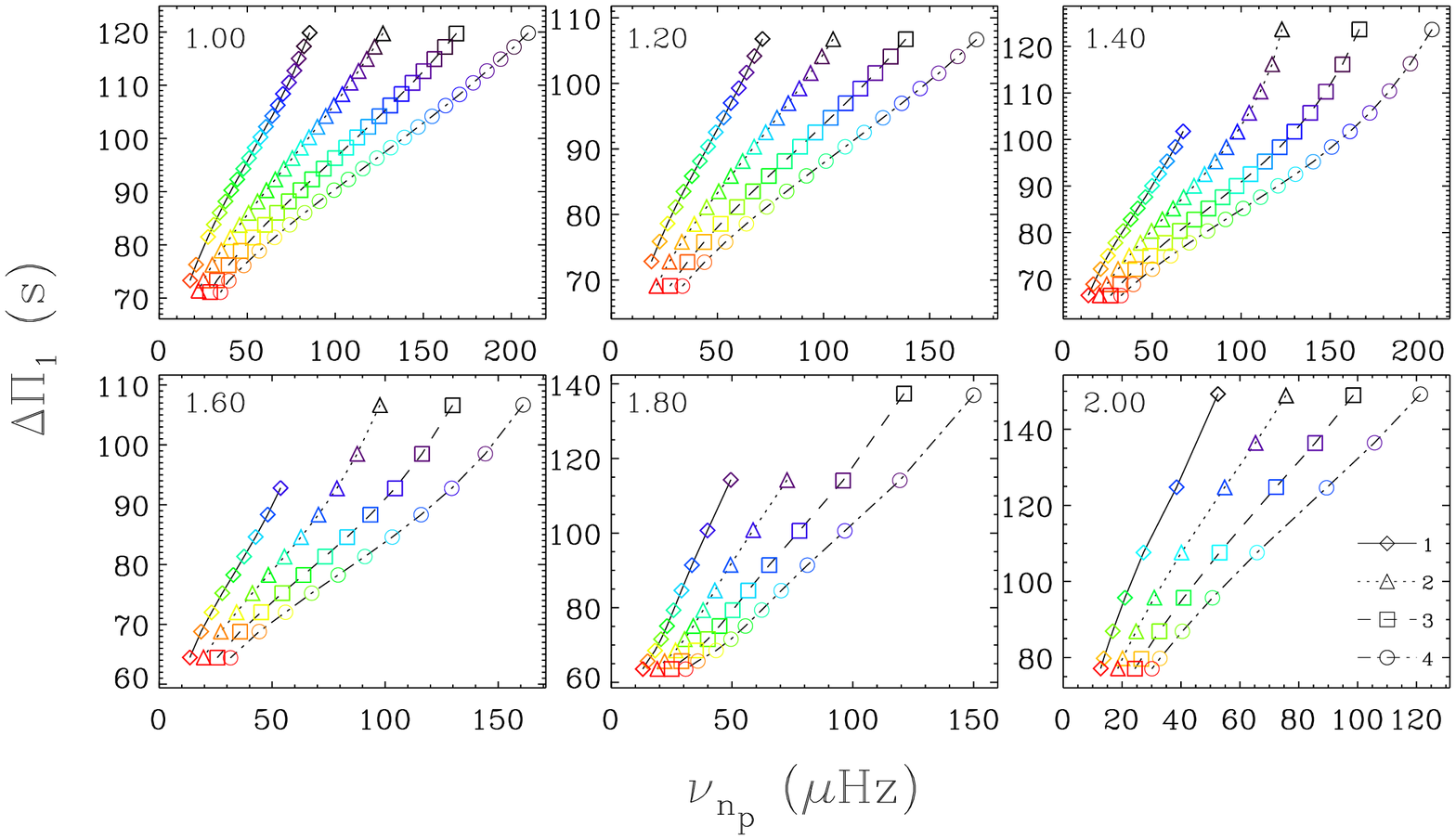}}
\caption{Variation of $\Dpi_1$ with stellar evolution for the low-$\np$ modes in models with $Z = 0.0295$. Colours and line notations are the same as in Figure~\ref{fg:low_qvar295}.} 
\label{fg:low_dpvar295}
\end{figure*}

\begin{figure*}
\resizebox{1.0\hsize}{!}{\includegraphics[angle =90]{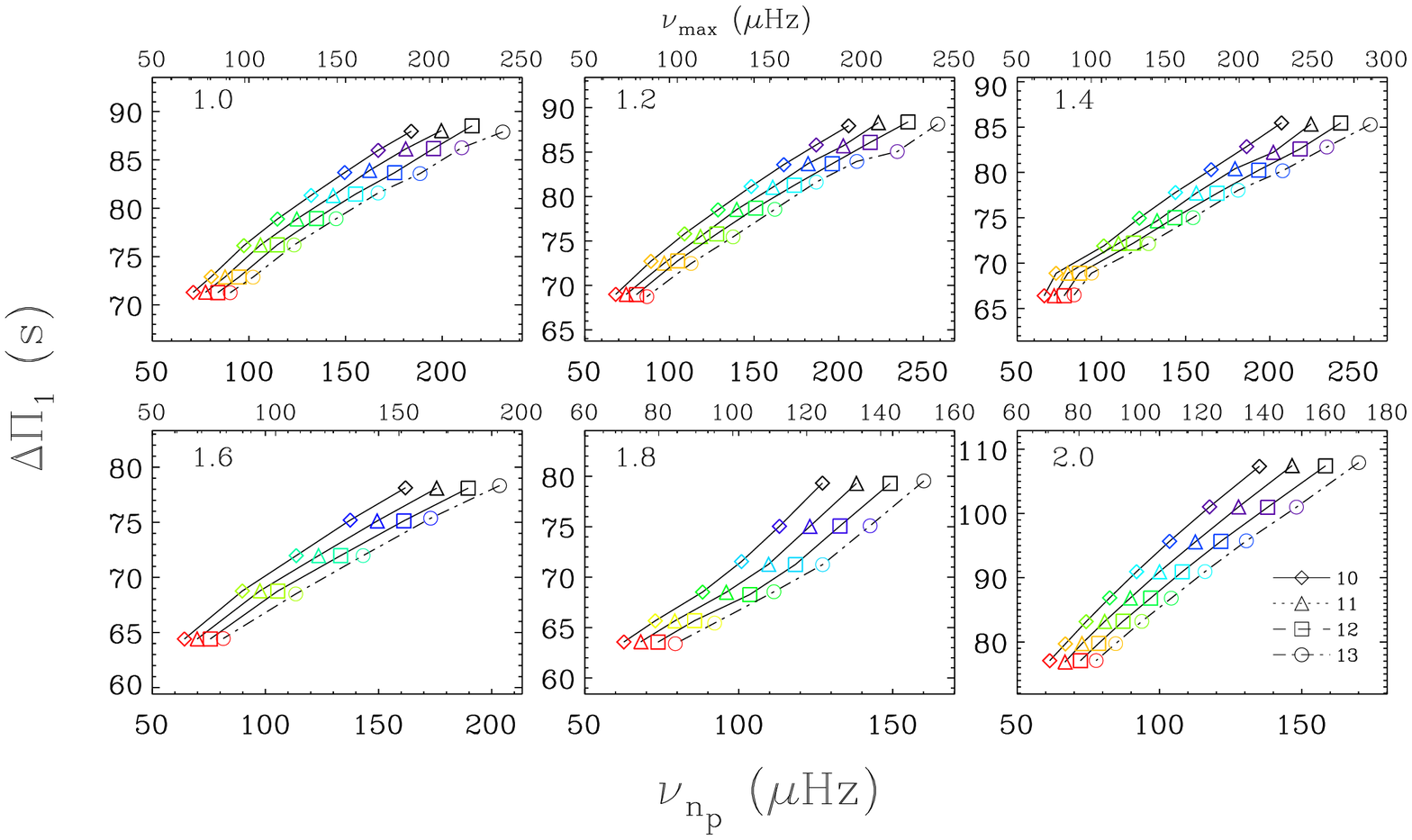}}
\caption{Variation of $\Dpi_1$ with stellar evolution for the high-$\np$ modes in models with $Z = 0.0295$. Colours and line notations are the same as in Figure~\ref{fg:high_qvar295}.} 
\label{fg:high_dpvar295}
\end{figure*}

From many previous studies such as \cite{mos12b}, we know that $\Dpi_1$ decreases when the star evolves on the RGB and hence it is an important seismic marker of stellar evolution. Evolution patterns of $\Dpi_1$ for the low-$\np$ modes are shown in Figure~\ref{fg:low_dpvar295}, in a format similar to the illustration in Figure~\ref{fg:low_qvar295} of the evolution of $q$. 
The decreasing trend is also $\np$-dependent, so that high-$\np$ modes vary more gradually with time. Therefore, we see an almost similar decreasing rate for the high-$\np$ modes show in Figure~\ref{fg:high_dpvar295}.
The illustration in Figure~\ref{fg:dpvarmass} of the dependence of the evolution of $\Dpi_1$ on mass shows that $\Dpi_1$ of the 1.80 and 2.00 $\msun$ models decreases much faster than that of models with smaller masses, indicating mass dependence of $\Dpi_1$ evolution. However, Figure~\ref{fg:dpvarz} shows almost uniform decreasing trends of $\Dpi_1$ for models with same mass but different $Z$, indicating no significant dependence on metallicity. 

A notable feature is that $\Dpi_1$ appears to be frequency independent for both low- and high-$\np$ modes, as seen from the fact that similar colours align horizontally in Figures~\ref{fg:low_dpvar295} and~\ref{fg:high_dpvar295}.
However, the asymptotic value of $\Dpi_1$ is related to the size of the g-mode cavity (cf. equation~\eqref{eq:dpi}). From Figure~\ref{fg:propagation} we can see that the g-mode cavity narrows down when the mode frequency increases, though the extent of the narrowing, and hence the change in $\Dpi_1$, is hard to determine from the figure. \cite{jiang14} analysed the character of the dipolar mixed modes with the Cowling approximation and indeed found that $\Dpi_1$ decreases as mode frequency increases. 
Here, we investigate oscillations based on the full set of equations without assuming the Cowling approximation.

\begin{figure}
\resizebox{1.0\hsize}{!}{\includegraphics[angle=90]{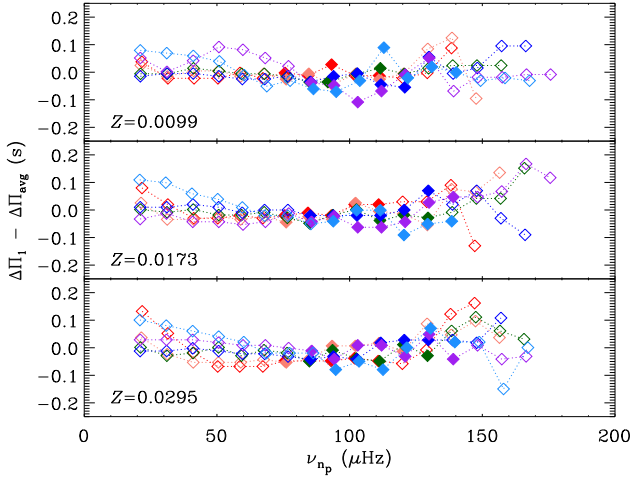}}
\caption{Variation of $\Dpi_1$ with frequency $\nu_{\np}$ for the $\Dnu \approx 9 \, \muHz$ models discussed in Section~\ref{sc:cpvarfre}. The variations are shown with respect to the average value $\Dpi_{\rm avg}$ computed from all the $\Dpi_1$ obtained for each model. The filled symbols show modes within the detectable solar-like frequency range.} 
\label{fg:dpglobal}
\end{figure}

By analysing the set of RGB models with $\Dnu$ around 9.0 $\muHz$ discussed in Section~\ref{sc:cpvar}, we are able to examine whether $\Dpi_1$ varies with frequency for high-$\np$ modes in evolved RGB models. The results are shown in Figure~\ref{fg:dpglobal}. The variations of the $\Dpi_1$ are given relative to the average $\Dpi_{\rm avg}$ computed from all the $\Dpi_1$ values estimated from each avoided crossing for a certain model. We use $\Dpi_{\rm avg}$ as a reference to produce variations around zero so that they can be compared across models, and to highlight the small amplitudes of the variations. 
The modes with highest frequencies are associated with high $\np$ but a low density of mixed modes. Hence the estimations of $\Dpi_1$ and $q$ have relatively large uncertainties. Therefore we see some fluctuations of $\Dpi_1$ at the high-frequency end. Other than those fluctuations, no significant variations of $\Dpi_1$ are found globally. The largest amplitudes of the variations are about 0.2~s which is tiny compared to the value of $\Dpi_1$ ($< 0.2 \%$) for these modes. Moreover, 0.2~s corresponds to the typical uncertainties in $\Dpi_1$ when this parameter is derived from fitting these low-density high-$\np$ modes. But the typical uncertainties in $\Dpi_1$ values inferred from the more stable fittings to mixed modes that are dense enough is around 0.05 s, which is also consistent with the amplitude of the period spacing variations found for those modes around $\numax$. Thus, the variations of $\Dpi_1$ seen in Figure~\ref{fg:dpglobal} stem mainly from the fitting.

We have investigated the variation of $\Dpi_1$ around $\Dpi_{\rm avg}$ in terms of the standard deviation of $\Dpi_1$ for modes with frequencies around $\numax$. 
We find the standard deviations to be less than 0.06 \% of the mean values. Thus, in the vicinity of $\numax$ where mixed modes can be observed, $\Dpi_1$ can be treated as a constant parameter when studying the character of mixed modes. This constancy of $\Dpi_1$ further motivates a comparison between the fitted and the asymptotic period spacing. To avoid the frequency dependence of the asymptotic period spacing, we define a proxy for this quantity, $\Dpi_{1, \rm bcz}$, by extending the integral in equation~\eqref{eq:dpi} to the full region between the centre and the base of the convection zone, defined by the radius "$r_{\rm bcz}$", such that
\begin{equation}
\Dpi_{1,{\rm bcz}} =  \frac{2\pi^2}{\sqrt{2}} \left(\int_{0}^{r_{\rm bcz}} N_{\rm BV} \frac{\mathrm dr}{r} \right)^{-1}.
\label{eq:dpi0}
\end{equation}
Thus, $\Dpi_{1,{\rm bcz}}$ is constant for a given model. The comparison between $\Dpi_1$ and $\Dpi_{1,{\rm bcz}}$ is shown in Figure~\ref{fg:dpdif} in terms of the absolute value of the difference between the two spacings relative to $\Dpi_1$. The mean relative difference is around $0.1\%$ and the largest difference, less than $1\%$, is found in the low g-mode density models that have larger uncertainties. Therefore, we find that $\Dpi_{1,{\rm bcz}}$ provides a good model representation of the constant $\Dpi_1$ derived directly from fitting the frequencies, in our RGB models.
 
\section{Buoyancy Glitch}
\label{sc:glitch}

As low-mass ( $\lesssim 2.2$ $\msun$) red-giant stars evolve on the RGB and reach the well-known luminosity bump, the increase in luminosity is interrupted when the hydrogen-burning shell approaches the composition discontinuity left behind by the retreating convective envelope during the first dredge-up. As a consequence of the decrease in the average mean molecular weight in the region just above the shell, the luminosity of the hydrogen-burning shell decreases, too. When the hydrogen-burning shell crosses the chemical discontinuity the luminosity resumes its increase. The amplitude of the luminosity variation in the bump is closely related to the magnitude of the hydrogen abundance difference at the discontinuity \citep{jcd15}. 

As a result of this chemical discontinuity, we find sharp variations in the buoyancy frequency, such as the spike signatures seen at the outer edge of the g-mode cavity for both RGB models in Figure~\ref{fg:propagation}. When the model approaches the luminosity bump, that rapid variation enters the g-mode propagation cavity. As the scale of variation of the chemical composition is comparable to or smaller than the local wavelength of the modes, it acts as a glitch in the background structure, hence, perturbing the phase of the wave, compared to that expected asymptotically from a smoothly varying background. The properties of the mixed modes in those RGB bump stars can be affected significantly in terms of their mode frequencies and inertias \citep{cunha15, cunha19}. These changes in the frequency pattern, in turn, can influence the inferred $\Dpi_1$ and $q$, if no account is taken of the presence of the glitch. According to \cite{cunha15}, for low-mass stars the glitches can be observed at the RGB luminosity bump, including a small region before the bump as well as the region where the luminosity decreases. In this section we seek to quantify the deviations in the inferred $q$ and $\Dpi_1$ parameters that are caused by the buoyancy glitch in RGB bump models.

Since equation~\eqref{eq:mixnu} is based on the asymptotic analysis, without any consideration of a buoyancy glitch, it is inadequate to estimate $q$ and $\Dpi_1$ for the red-giant bump stars. However, this limitation can somehow help us identify that a glitch is present and also quantify the impact of the glitch, as the values of $q$ and $\Dpi_1$ estimated from equation~\eqref{eq:mixnu} can be abnormal for models in the vicinity of the bump.

We select 10 models around the luminosity bump along our 1.0 $\msun$, solar metallicity evolutionary track (not included in the coloured section in Figure~\ref{fg:tracks}) for a brief analysis of the effect of the buoyancy glitch on the mixed modes in these bump models. Figure~\ref{fg:glitch}(a) shows the location of the models relative to the luminosity variation through the bump. The models are on the same evolutionary sequence as those presented in Figure~\ref{fg:propagation}. 
The youngest four models in darker colours are located before the bump. The following three models are in the stage where the luminosity temporarily decreases. 
The hydrogen-burning shell of these seven models still approaches the spike signature seen in Figure~\ref{fg:propagation} as the model evolves.
The remaining three models are located after the bump, where the luminosity resumes the increase. For these models the spike is no longer present as it has eventually been reached by the hydrogen-burning shell.

\begin{figure}
\resizebox{1.0\hsize}{!}{\includegraphics[angle=90]{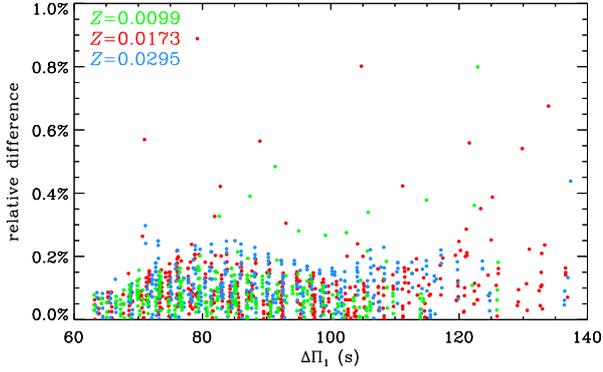}}
\caption{Relative difference between the fitted and the asymptotic spacings, defined as $|\Dpi_{1,{\rm bcz}} - \Dpi_1 | / \Dpi_1$.} 
\label{fg:dpdif}
\end{figure}

\begin{figure}
\resizebox{1.0\hsize}{!}{\includegraphics{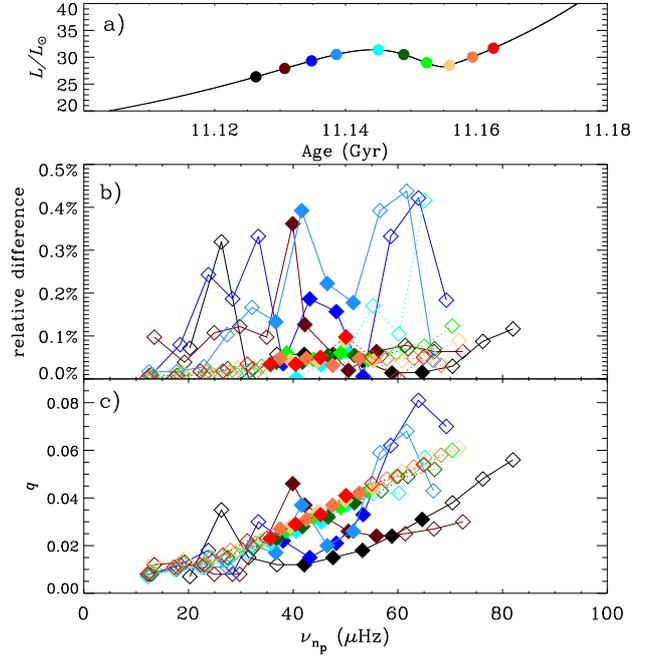}}
\caption{(a) Evolution near the RGB luminosity bump exhibiting the luminosity as a function of age for the $1.0$ $\msun$, solar metallicity model. The positions of the 10 selected models are highlighted by the coloured dots. (b)--(c) Relative difference of period spacings, defined in Figure~\ref{fg:dpdif}, and $q$ for the 10 models, as a function of $\nu_{\np}$. Modes with the same age are connected by the same line (before bump: solid line, other: dotted line). The symbols use the same colouring as in panel (a). Modes around $\numax$ are indicated by filled symbols.} 
\label{fg:glitch}
\end{figure}

Figure~\ref{fg:glitch}(b)--(c) show the
frequency behaviour of the relative period-spacing differences defined in Figure~\ref{fg:dpdif}, as well as the frequency dependence of $q$ for the dipolar mixed modes in the 10 models. The $\Dpi_1$ and $q$ are estimated from equation~\eqref{eq:mixnu}, using the method described in Section~\ref{sc:fittingMethod}. Fluctuations in the period spacing difference are seen in the youngest four models located before the bump. More pronounced fluctuations are also seen in $q$ for the same four models. For the remaining 6 older models, the relative differences of the period spacings (defined in the caption of Figure~\ref{fg:dpdif}) show little variation with frequency, which is consistent with the conclusion of Section~\ref{sc:psvar}. The $q$ values are also remarkably similar and follow the same increasing trend with $\nu_{\np}$, which is clearly different from the behaviour of $q$ in the first four models, before the bump. Thus we find more significant glitch-induced variations in models before the bump, when the luminosity still increases. 
In fact, while the spike is still present in the buoyancy frequency of the models in the region of decreasing luminosity, it has little impact on the oscillations. This is because the scale of the spike becomes larger than the local wavelength in these models. In addition to that, the position of the spike relative to the middle of the g-mode cavity is changing rapidly, which largely reduces the impact on the inferred $\Dpi_1$ \citep{cunha15}.

It should also be mentioned that the uniform increase of $q$ for this set of more evolved models confirms that $q$ increases with mode frequency, and the nearly coincident increasing trends indicate that $q$ varies slowly with stellar evolution for high-$\np$ modes, as concluded in Section~\ref{sc:cpvar}.

This brief analysis tells us that the buoyancy glitch results in significant variations of the inferred $q$ and $\Dpi_1$ if they are derived without including the glitch effect. The impact vanishes as the shell crosses the discontinuity. A more comprehensive analysis has been performed to characterize the buoyancy glitches for RGB bump models by using the analytical expression for the period spacing proposed in \cite{cunha19}.

We also perform a linear fit to the modes around $\numax$ for the 6 oldest models, using the method mentioned in Section~\ref{sc:cpvarhighfre}. The results are given in Table~\ref{tb:qvarglitch}, from which we can see that for these more evolved models the slopes $B$ are much larger than what is given in Table~\ref{tb:qvarfit}, confirming that the increase of $q$ with frequency becomes faster as the star evolves on the RGB.

\section{Conclusion}

Hundreds of RGB models with different masses and metallicities have been analysed to study the mixing behaviour of dipolar mixed modes in red-giant stars. The oscillation modes are fitted by the Bayesian-based model selection program {\scriptsize DIAMONDS} to obtain the g-mode period spacing and the coupling strength. The coupling strength decreases as the model climbs up the RGB and the slope of the decreasing trend is dependent on the radial order of the pressure mode component $\np$. For low-$\np$ modes the coupling strength decreases faster as the star evolves, while for high-$\np$ modes, such as those with frequencies in the observable frequency range, the decrease in coupling strength becomes very gradual. 
The rate at which the coupling strength decreases is larger for higher-mass and metal-poorer models. Similar evolutionarily decreasing behaviour is found for the period spacing. While its decrease depends on model mass, no significant dependence on metallicity is shown.

With the analysis of evolved RGB models that have $\Dnu \approx 9.0\,\muHz$ we notice a non-negligible increase of the coupling strength as the mode frequency increases. The increase becomes faster for more evolved models. Hence, a constant coupling strength should be used with caution when studying the mixed modes of evolved RGB stars. 
However, abnormal values of the coupling strength are also seen for low-frequency modes, which need further investigation. 
On the other hand, no noticeable change of the period spacing with frequency is found in our models (with the exception of the models immediately before the luminosity bump); in particular, in the observable frequency range the period spacing can be regarded as a constant parameter. 
Moreover, we showed that its value is consistent with the asymptotic one computed when the integral of the buoyancy frequency is considered between the centre and the base of the convection zone.
For the models located right before the luminosity bump on the RGB, the frequency behaviour of the coupling strength and period spacing is different. Here we find that both parameters show significant fluctuations in frequency which are not real, but rather result from equation~\eqref{eq:mixnu} being inadequate to describe the frequencies in models where a glitch in present. 

This analysis opens the way to explore further the influence of stellar parameters on the mixing character of the oscillation modes. Such study will be conducted with a larger set of models. To further investigate the differences between the asymptotic parameters and the results obtained from fitting the mode frequencies, a detailed examination of the models' inner structure should also be considered in future work.
Here a comparison with the analytical results obtained by \citet{pin20} will be very interesting.

\section{Acknowledgements}

The authors wish to thank the referee for pertinent comments on the manuscript that greatly improved the presentation. We acknowledge support from the National Key Program for Science and Technology Research and Development (2017YFB0203300). Funding for the Stellar Astrophysics Centre is provided by The Danish National Research Foundation (Grant DNRF106). M. S. Cunha is supported by national funds through FCT in the form of a work contract. This work was supported by FCT through national funds (PIDDAC) (grants: PTDC/FIS-AST/30389/2017 and UID/FIS/04434/2019) and by FEDER - Fundo Europeu de Desenvolvimento Regional through COMPETE2020 - Programa Operacional Competitividade e Internacionalização (grant: POCI-01-0145-FEDER-030389). This work is also funded by the Fundamental Research Funds for the Central Universities (grant: 19lgpy278). Funding for Yunnan Observatories is cosponsored by the National Natural Science Foundation of China (grant No. 11303087), Ten Thousands Talents Program of Yunnan Province, foundations of the Chinese Academy of Sciences (Light of West China Program, Youth Innovation Promotion Association, and the B-type Strategic Priority Program Grant No. XDB41000000).

\clearpage

\begin{table}
\caption{Parameters from the linear fit of equation~\eqref{eq:qlinfit} to the frequencies of modes around $\numax$ shown in Figure~\ref{fg:qdif295} ($q_1, \alpha$). The pair of parameters ($A,B$) is derived from ($q_1, \alpha$) using equations~\eqref{eq:a} and~\eqref{eq:b}.}
 \label{tb:qvarfit} 
 \centering
\begin{tabular}{@{}lccccccc}
\hline
\small{$M$} & $Z$ & $\numax$ & $q_1$ & $\alpha_1$ & $B$ & $A$ \\
\small $(\msun)$ & & $(\muHz)$ & & & ($\times 10^{-3} \, \muHz^{-1}$) \\
\hline
\small 1.00 & \small 0.0099 & \small 92.50 & \small $0.114 \pm 0.015$ & $\small 0.626 \pm 0.067$ & $\small 0.772 \pm 0.131$ & $\small 0.043 \pm 0.009$\\
\small 1.20 & \small 0.0099 & \small 97.66 & \small $0.108 \pm 0.012$ & $\small 0.552 \pm 0.057$ & $\small 0.610 \pm 0.093$ & $\small 0.048 \pm 0.008$\\
\small 1.40 & \small 0.0099 & \small 102.38 & \small $0.100 \pm 0.015$ & $\small 0.450 \pm 0.074$ & $\small 0.440 \pm 0.098$ & $\small 0.055 \pm 0.011$\\
\small 1.60 & \small 0.0099 & \small 106.51 & \small $0.095 \pm 0.008$ & $\small 0.647 \pm 0.046$ & $\small 0.577 \pm 0.064$ & $\small 0.034 \pm 0.005$ \\
\small 1.80 & \small 0.0099 & \small 110.30 & \small $0.117 \pm 0.017$ & $\small 1.154 \pm 0.120$ & $\small 1.22 \pm 0.129$ & $\small -0.018 \pm 0.014$ \\
\small 2.00 & \small 0.0099 & \small 114.27 & \small $0.141 \pm 0.008$ & $\small 0.555 \pm 0.030$ & $\small 0.685 \pm 0.037$ & $\small 0.063 \pm 0.006$ \\
\hline
\small 1.00 & \small 0.0173 & \small 93.73 & \small $0.105 \pm 0.014$ & $\small 0.607 \pm 0.069$ & $\small 0.680 \pm 0.078$ & $\small 0.041 \pm 0.009$\\
\small 1.20 & \small 0.0173 & \small 98.96 & \small $0.100 \pm 0.010$ & $\small 0.505 \pm 0.052$ & $\small 0.510 \pm 0.053$ & $\small 0.050 \pm 0.007$ \\
\small 1.40 & \small 0.0173 & \small 103.66 & \small $0.093 \pm 0.011$ & $\small 0.673 \pm 0.065$ & $\small 0.604 \pm 0.059$ & $\small 0.030 \pm 0.007$\\
\small 1.60 & \small 0.0173 & \small 107.80 & \small $0.087 \pm 0.013$ & $\small 0.787 \pm 0.086$ & $\small 0.635 \pm 0.070$ & $\small 0.019 \pm 0.008$ \\
\small 1.80 & \small 0.0173 & \small 111.74 & \small $0.081 \pm 0.007$ & $\small 0.947 \pm 0.062$ & $\small 0.686 \pm 0.045$ & $\small 0.004 \pm 0.005$ \\
\small 2.00 & \small 0.0173 & \small 114.18 & \small $0.131 \pm 0.010$ & $\small 0.617 \pm 0.039$ & $\small 0.708 \pm 0.045$ & $\small 0.050 \pm 0.006$ \\
\hline
\small 1.00 & \small 0.0295 & \small 94.76 & \small $0.101 \pm 0.010$ & $\small 0.612 \pm 0.053$ & $\small 0.652 \pm 0.057$ & $\small 0.039 \pm 0.007$ \\
\small 1.20 & \small 0.0295 & \small 100.07 & \small $0.093 \pm 0.013$ & $\small 0.463 \pm 0.071$ & $\small 0.430 \pm 0.067$ & $\small 0.050 \pm 0.010$ \\
\small 1.40 & \small 0.0295 & \small 104.70 & \small $0.085 \pm 0.009$ & $\small 0.755 \pm 0.064$ & $\small 0.611 \pm 0.052$ & $\small 0.021 \pm 0.006$ \\
\small 1.60 & \small 0.0295 & \small 108.95 & \small $0.083 \pm 0.011$ & $\small 0.809 \pm 0.084$ & $\small 0.616 \pm 0.065$ & $\small 0.016 \pm 0.007$ \\
\small 1.80 & \small 0.0295 & \small 112.81 & \small $0.083 \pm 0.007$ & $\small 0.777 \pm 0.049$ & $\small 0.572 \pm 0.036$  & $\small 0.019 \pm 0.004$\\
\small 2.00 & \small 0.0295 & \small 116.62 & \small $0.118 \pm 0.014$ & $\small 0.703 \pm 0.066$ & $\small 0.711 \pm 0.067$ & $\small 0.035 \pm 0.009$ \\
\hline
\end{tabular}
\end{table}

\begin{table}
\caption{Mean values of the pairs ($A, B$) given in Table~\ref{tb:qvarfit} for models with same $Z$ but different masses.}
 \label{tb:qvarZ} 
 \centering
 \begin{threeparttable}
\begin{tabular}{@{}lccc}
\hline
\small{Z} & \small 0.0099 \tnote{a} & \small 0.0173 & \small 0.0295 \\
\hline
\small{$A$} & $\small 0.049 \pm 0.004$ & $\small 0.032 \pm 0.003$ & $\small 0.030 \pm 0.003$ \\
\small{$B (\times 10^{-3} \, \muHz^{-1})$} &$\small 0.617 \pm 0.040$ & $\small 0.637 \pm 0.024$ & $\small 0.599 \pm 0.024$\\
\hline
\end{tabular}
\begin{tablenotes}
\item [a] {\scriptsize 1.80 $\msun$ model is not taken into account.}
\end{tablenotes}
\end{threeparttable}
\end{table}

\begin{table}
\caption{Mean values of the pairs ($A, B$) given in Table~\ref{tb:qvarfit} for models with same mass but different $Z$.}
 \label{tb:qvarM} 
 \centering
  \begin{threeparttable}
\begin{tabular}{@{}lccccccc}
\hline
\small{$M (\msun)$} & \small 1.00 & \small 1.20 & \small 1.40 &\small 1.60 & \small 1.80\tnote{a} & \small 2.00 \\
\hline
\small{$A$} & $ \small 0.41\pm 0.005$ & $\small 0.049 \pm 0.005$ & $\small  0.035 \pm 0.005$ & $\small 0.023 \pm 0.004$ & $ \small 0.012 \pm 0.003$ & $\small 0.049 \pm 0.006$\\
\small{$B (\times 10^{-3} \, \muHz^{-1})$} & $\small 0.701 \pm 0.054$ & $ \small 0.517 \pm 0.042$ & $\small 0.626 \pm 0.042$ & $\small 0.609 \pm 0.038$& $\small 0.629 \pm 0.029$& $\small 0.701 \pm 0.044$\\
\hline
\end{tabular}
\begin{tablenotes}
\item [a] {\scriptsize $Z = 0.0099$ model is not taken into account.}
\end{tablenotes}
\end{threeparttable}\end{table}

\begin{table}
\caption{Similar to Table~\ref{tb:qvarfit}, parameters from the linear fit of equation~\eqref{eq:qlinfit} to the frequencies of modes around $\numax$ in the 6 oldest models discussed in Section~\ref{sc:glitch}.}
 \label{tb:qvarglitch} 
 \centering
\begin{tabular}{@{}lcccccc}
\hline
\small{$L$} & $\numax$ & $q_1$ & $\alpha_1$ & $B$ & $A$ \\
\small $(L_{\sun})$  & $(\muHz)$ & & & ($\times 10^{-3} \, \muHz^{-1}$) \\
\hline
\small 31.378 &  \small 42.78 & \small $0.028 \pm 0.005$ & $\small 1.633 \pm 0.228$ & $\small 1.069 \pm 0.152$ & $ \small -0.018 \pm 0.007$ \\
\small 30.498 & \small 44.32 & \small $0.030 \pm 0.003$ & $\small 1.446 \pm 0.113$ & $\small 0.979 \pm 0.077$ & $\small -0.013 \pm 0.004$ \\
\small 28.982 &  \small 47.17 & \small $0.035 \pm 0.005$ & $\small 1.431 \pm 0.147$ & $\small 1.062 \pm 0.110$ & $\small -0.015 \pm 0.006$ \\
\small 28.469 &  \small 48.24 & \small $0.037 \pm 0.002$ & $\small 1.304 \pm 0.060$ & $\small 1.000 \pm 0.046$ & $\small -0.011 \pm 0.002$ \\
\small 30.028 &  \small 45.30 & \small $0.034 \pm 0.003$ & $\small 1.329 \pm 0.094$ & $\small 0.997 \pm 0.071$ & $ \small -0.011 \pm 0.003$ \\
\small 31.692 &  \small 42.49 & \small $0.031 \pm 0.006$ & $\small 1.651 \pm 0.263$ & $\small 1.205 \pm 0.195$ & $ \small -0.020 \pm 0.009$ \\
\hline
\end{tabular}
\end{table}

\clearpage

\begin{figure*}
\resizebox{1.0\hsize}{!}{\includegraphics[angle =0]{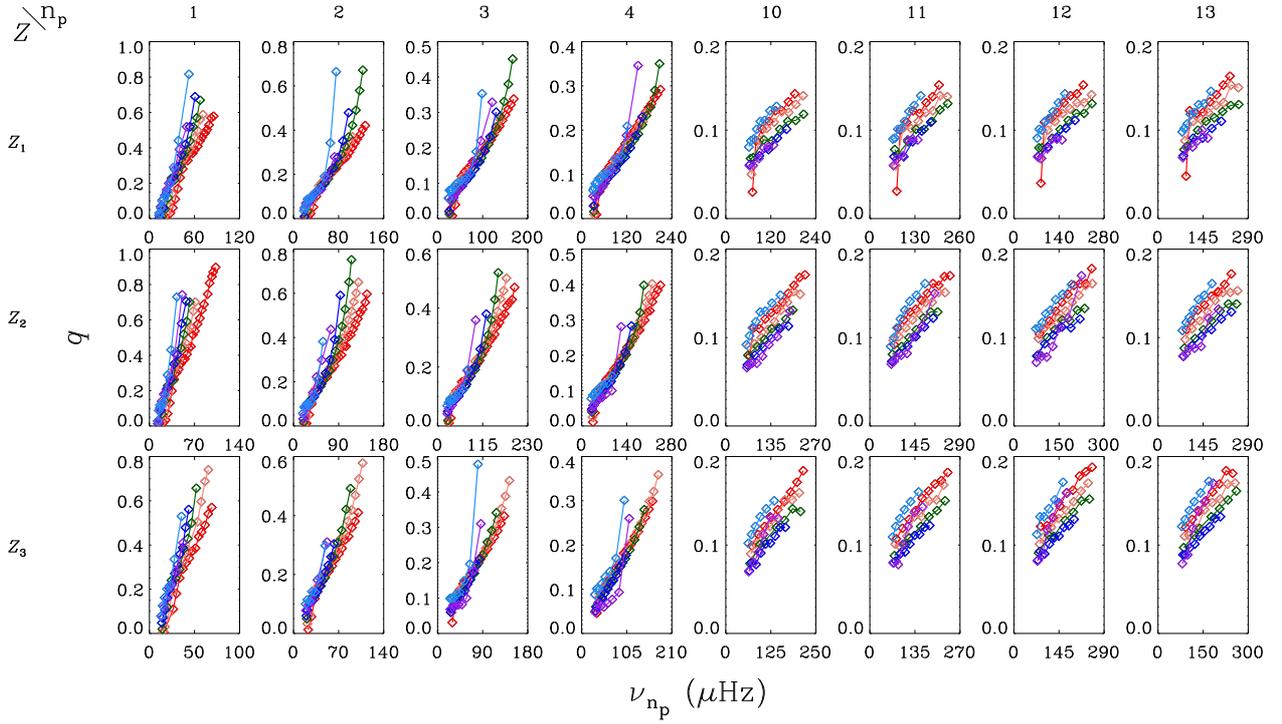}}
\caption{Array of diagrams showing the mass dependence of the evolution of $q$ with age, reflected in the variation of $\nu_{\np}$. Diagrams in the same row share the same initial chemical abundance (given on the left of the figure, $Z_1: 0.0295,\,Z_2: 0.0173,\,Z_3: 0.0099$), while the ones in the same column share a same $\np$ (values given on the top). The colour of the symbols indicates the model mass, using the same colouring as in Figure~\ref{fg:tracks} (red: 1.00 $\msun$, orange: 1.20 $\msun$, green: 1.40 $\msun$, blue: 1.6 $\msun$, purple: 1.8 $\msun$, cyan: 2.00 $\msun$). Model age is not given, but it is indicated by $\nu_{\np}$ with younger models having larger $\nu_{\np}$ values. Note the different scale of $q$ in each diagram.} 
\label{fg:qvarmass}
\end{figure*} 

\begin{figure*}
\resizebox{1.0\hsize}{!}{\includegraphics[angle = 0]{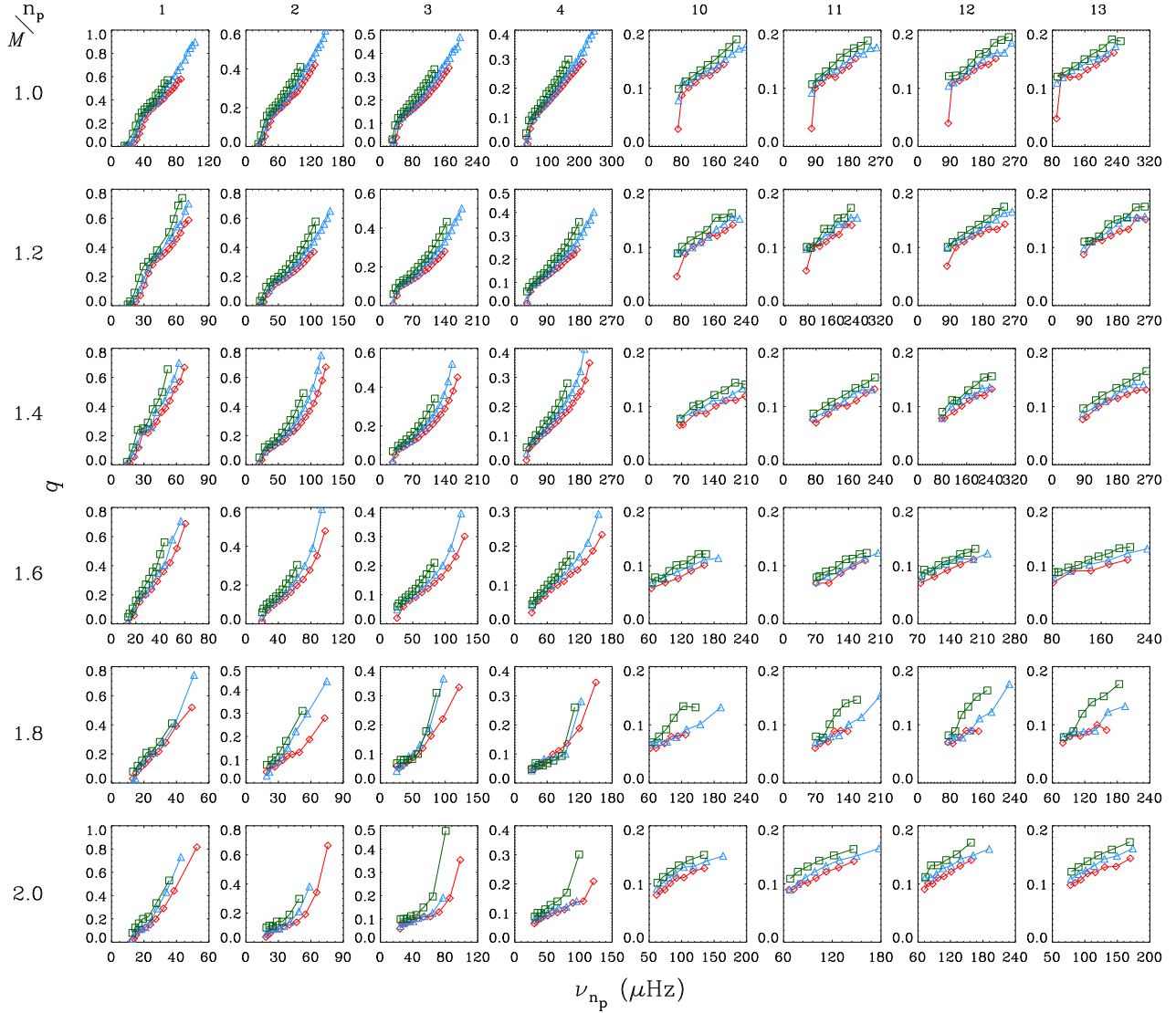}}
	\caption{Array of diagrams showing the metallicity dependence of the evolution of $q$ with age, reflected in the variation of $\nu_{\np}$ similar to Figure~\ref{fg:qvarmass}, but with diagrams in the same column sharing the same initial mass (values given at the top of the figure in units of solar mass). The different metallicities of $Z = 0.0099$ (green squares), 0.0173 (orange triangles) and 0.0295 (red diamonds) clearly have an effect on the decrease of $q$. Note the different scale of $q$ in each diagram.} 
\label{fg:qvarz}
\end{figure*} 

\begin{figure*}
\resizebox{1.0\hsize}{!}{\includegraphics[angle =0]{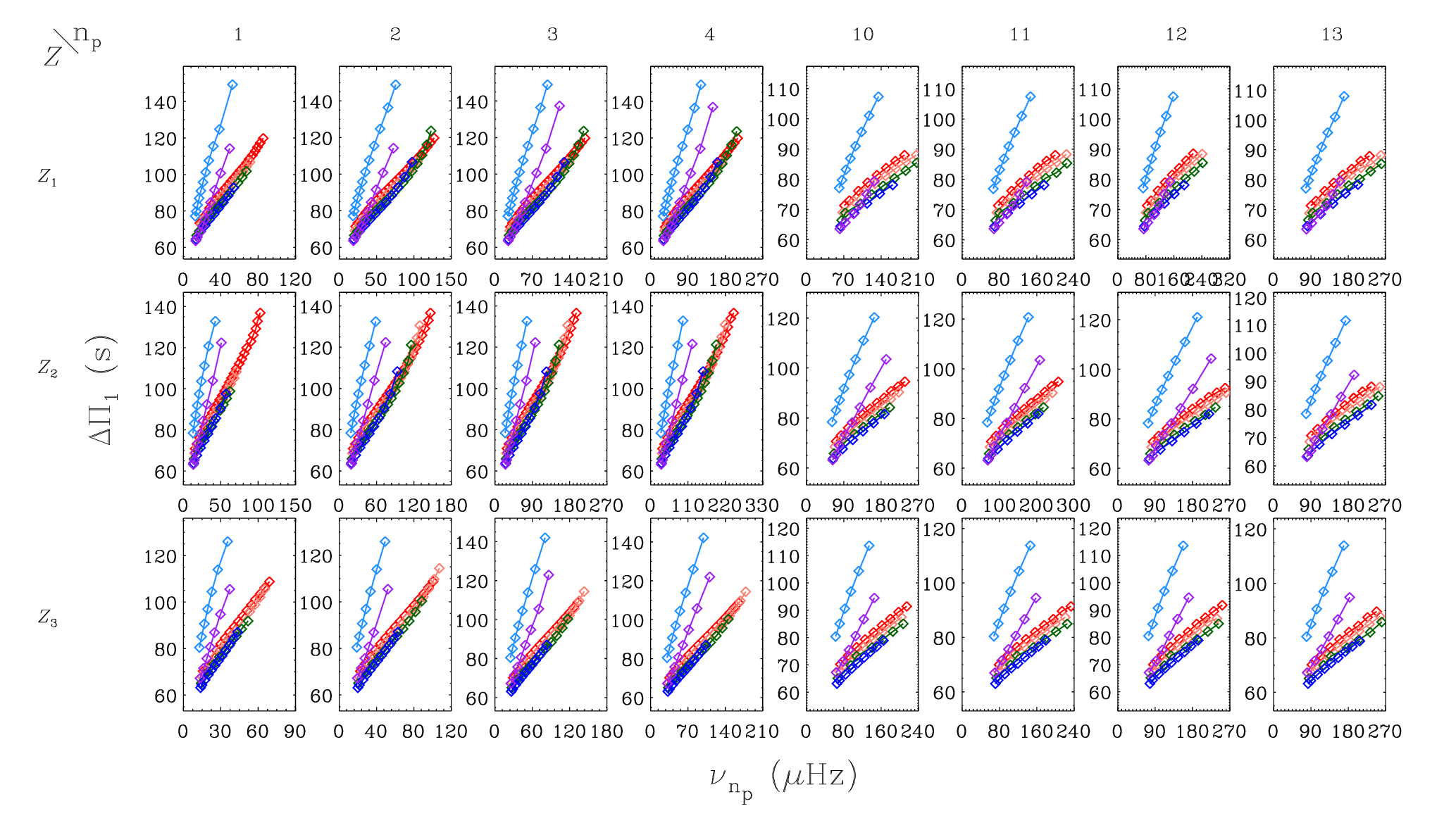}}
\caption{Array of diagrams showing the mass dependence of the evolution of $\Dpi_1$. Plot settings refer to Figure~\ref{fg:qvarmass}.} 
\label{fg:dpvarmass}
\end{figure*} 

\begin{figure*}
\resizebox{1.0\hsize}{!}{\includegraphics[angle =0]{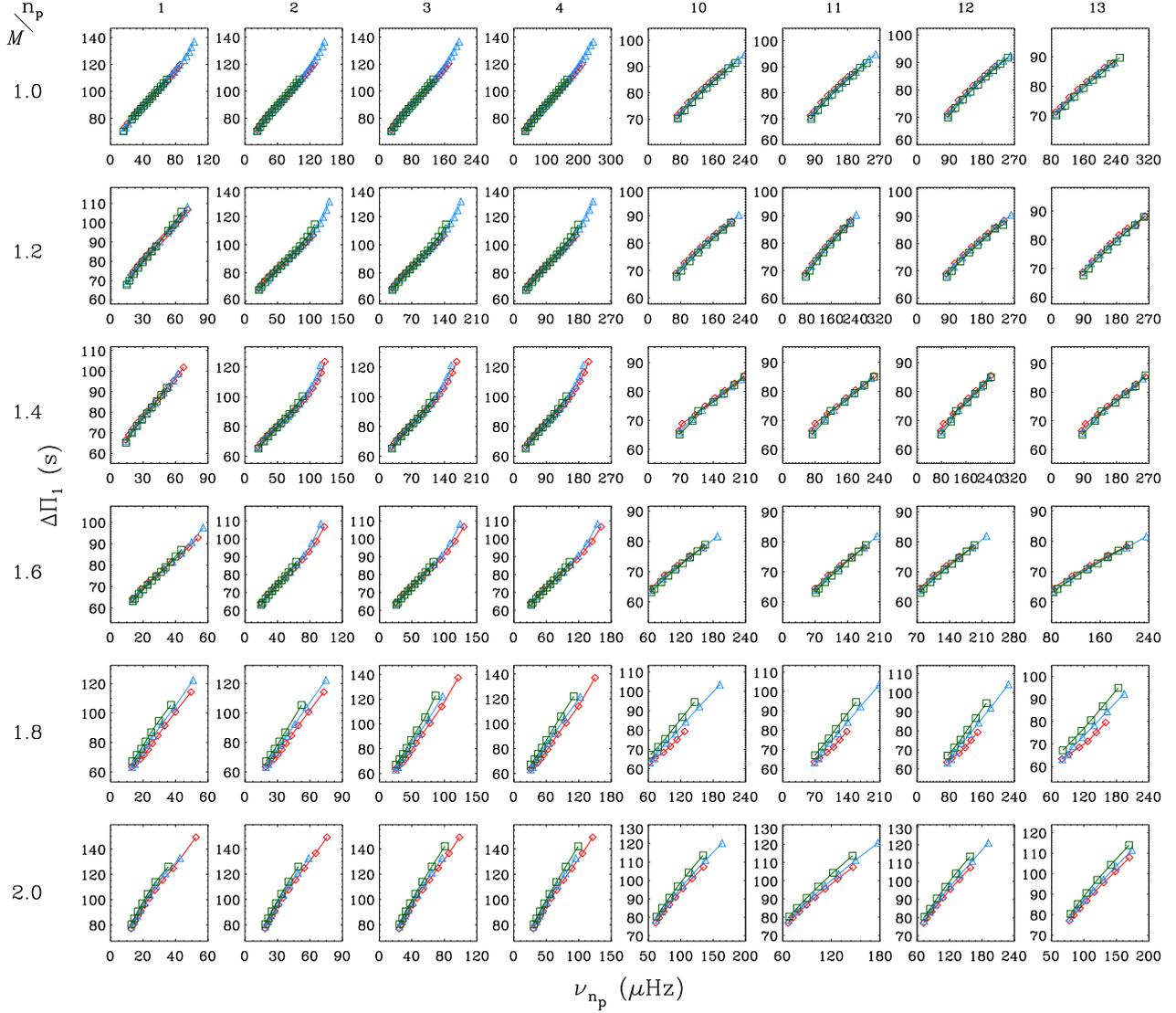}}
\caption{Array of diagrams showing the metallicity dependence of the evolution of $\Dpi_1$. Plot settings refer to Figure~\ref{fg:qvarz}.} 
\label{fg:dpvarz}
\end{figure*} 

\label{lastpage}


\begin{thebibliography}{50}

\bibitem[\protect\citeauthoryear{Aerts et al.}{2010}]{bookas} Aerts, C.,
Christensen-Dalsgaard, J.,
\& Kurtz, D.~W.\ 2010, Asteroseismology by C.~Aerts, J.~Christensen-Dalsgaard, and D.W.~Kurtz.~Springer, 2010

\bibitem[Aizenman et 
al.(1977)]{aiz77} Aizenman, M., Smeyers, P., \& Weigert, A.\ 1977, A\&A, 58, 41 

\bibitem[Angulo et al.(1999)]{ang99} Angulo, C., et al.\
1999, Nuclear Physics A, 656, 3

\bibitem[\protect\citeauthoryear{Baudin et 
al.}{2012}]{bau12} Baudin F., et al., 2012, A\&A, 538, A73 

\bibitem[\protect\citeauthoryear{Beck et al.}{2011}]{bec11} Beck, P.~G., Bedding, 
T.~R., Mosser, B., et al. 2011, Science, 332, 205 

\bibitem[\protect\citeauthoryear{Beck et al.}{2012}]{bec12} Beck, P.~G., Montalban, 
J., Kallinger, T., et al. 2012, Nature, 481, 55 

\bibitem[Bedding et al.(2007)]{bed07} Bedding, T.~R., 
Kjeldsen, H., Arentoft, T., et al.\ 2007, ApJ, 663, 1315

\bibitem[\protect\citeauthoryear{Bedding et al.}{2010}]{bed10} Bedding, T.~R., Huber, 
D., Stello, D., et al. 2010, ApJ, 713, L176

\bibitem[\protect\citeauthoryear{Bedding et 
al.}{2011}]{bed11} Bedding T.~R., et al., 2011, Nature, 471, 
608

\bibitem[Bedding(2012)]{bed12} Bedding, T.~R.\ 2012, Progress 
in Solar/Stellar Physics with Helio- and Asteroseismology, 462, 195

\bibitem[\protect\citeauthoryear{Benomar et al.}{2012}]{ben12} Benomar, O., Bedding, 
T.~R., Stello, D., et al. 2012, ApJ, 745, L33

\bibitem[B{\"o}hm-Vitense(1958)]{boh58} B{\"o}hm-Vitense, E. 1958, Z. Astrophys., 46, 108

\bibitem[Brand{\~a}o et 
al.(2011)]{bra11} Brand{\~a}o, I.~M., Do{\u g}an, G., Christensen-Dalsgaard, J., et al.\ 2011, A\&A, 527, A37

\bibitem[Buysschaert et al.(2016)]{buy16} Buysschaert, B., Beck, P.~G., Corsaro, E., et al.\ 2016, \aap, 588, A82

\bibitem[Carrier et 
al.(2005)]{car05} Carrier, F., Eggenberger, P., \& Bouchy, F.\ 2005, A\&A, 434, 1085

\bibitem[Christensen-Dalsgaard et al.(1995)]{jcd95} 
Christensen-Dalsgaard, J., Bedding, T.~R., 
\& Kjeldsen, H.\ 1995, ApJ, 443, L29

\bibitem[\protect\citeauthoryear{Christensen-Dalsgaard}{2008a}]{jcd08a} Christensen-Dalsgaard, J. 2008a, Ap\&SS, 316, 13

\bibitem[\protect\citeauthoryear{Christensen-Dalsgaard}{2008b}]{jcd08b} Christensen-Dalsgaard, J.\ 2008b, Ap\&SS, 316, 113 

\bibitem[\protect\citeauthoryear{Christensen-Dalsgaard}{2012}]{jcd12} 
Christensen-Dalsgaard, J. 2012, Progress in Solar/Stellar Physics with 
Helio- and Asteroseismology, 462, 503

\bibitem[Christensen-Dalsgaard(2015)]{jcd15} Christensen-Dalsgaard, J.\ 2015, \mnras, 453, 666 

\bibitem[Corsaro \& De Ridder(2014)]{cor14} Corsaro, E., \& De Ridder, J.\ 2014, A\&A, 571, A71 

\bibitem[Cowling(1941)]{cow41} Cowling, T.~G.\ 1941, \mnras, 101, 367

\bibitem[Cunha et al.(2015)]{cunha15} Cunha, M.~S., Stello, D., Avelino, P.~P., Christensen-Dalsgaard, J., \& Townsend, R.~H.~D.\ 2015, \apj, 805, 127 

\bibitem[Cunha et al.(2019)]{cunha19} Cunha, M.~S., Avelino, P.~P., Christensen-Dalsgaard, J., et al.\ 2019, \mnras, 490, 909

\bibitem[Deheuvels 
\& Michel(2010)]{deh10} Deheuvels, S., \& Michel, E.\ 2010, Ap\&SS, 328, 259

\bibitem[\protect\citeauthoryear{Gough}{1993}]{gou93} Gough, D.~O. 1993, 
Astrophysical Fluid Dynamics - Les Houches 1987, 399

\bibitem[Grec et al.(1983)]{gre83} Grec, G., Fossat, E., 
\& Pomerantz, M.~A.\ 1983, SoPh, 82, 55

\bibitem[\protect\citeauthoryear{Hekker et 
al.}{2009}]{hek09} Hekker, S., et al., 2009, A\&A, 506, 465

\bibitem[Hekker \& Christensen-Dalsgaard(2017)]{hek17} Hekker, S., \& Christensen-Dalsgaard, J.\ 2017, \aapr, 25, 1

\bibitem[Hekker et al.(2018)]{hek18} Hekker, S., Elsworth, Y., \& Angelou, G.~C.\ 2018, \aap, 610, A80 

\bibitem[\protect\citeauthoryear{Huber et al.}{2010}]{hub10} Huber, D., et al., 2010, ApJ, 723, 1607 

\bibitem[Iglesias \& Rogers(1996)]{igl96} Iglesias, C.~A., \& Rogers, F.~J.\ 1996, ApJ, 464, 943

\bibitem[\protect\citeauthoryear{Jiang et al.}{2011}]{jiang11} 
Jiang C., et al., 2011, ApJ, 742, 120

\bibitem[\protect\citeauthoryear{Jiang 
\& Christensen-Dalsgaard}{2014}]{jiang14} Jiang C., Christensen-Dalsgaard J., 2014, MNRAS, 444, 3622

\bibitem[Jiang et al.(2018)]{jiang18} Jiang, C., Christensen-Dalsgaard, J., \& Cunha, M.\ 2018, \mnras, 474, 5413

\bibitem[\protect\citeauthoryear{Kallinger et 
al.}{2012}]{kal12} Kallinger, T., et al., 2012, A\&A, 541, A51

\bibitem[Kjeldsen et al.(1995a)]{hans95a} Kjeldsen, H., Bedding, 
T.~R., Viskum, M., \& Frandsen, S.\ 1995, AJ, 109, 1313

\bibitem[Kjeldsen \& Bedding(1995b)]{hans95b} Kjeldsen, H., \& Bedding, T.~R.\ 1995, \aap, 293, 87

\bibitem[Kjeldsen et al.(2003)]{hans03} Kjeldsen, H., Bedding, 
T.~R., Baldry, I.~K., et al.\ 2003, AJ, 126, 1483

\bibitem[\protect\citeauthoryear{Kjeldsen, Bedding, 
\& Christensen-Dalsgaard}{2008}]{hans08} Kjeldsen H., Bedding T.~R., Christensen-Dalsgaard J., 2008, ApJ, 683, L175

\bibitem[\protect\citeauthoryear{Mathur et al.}{2011}]{mat11} Mathur, S., et al., 2011, ApJ, 733, 95

\bibitem[Montalb{\'a}n et al.(2013)]{mon13} Montalb{\'a}n, J., Miglio, A., Noels, A., et al.\ 2013, \apj, 766, 118

\bibitem[\protect\citeauthoryear{Mosser et 
al.}{2011}]{mos11} Mosser, B., et al., 2011, A\&A, 532, A86

\bibitem[Mosser et al.(2012a)]{mos12a} Mosser, B., Goupil, M.~J., Belkacem, K., et al.\ 2012a, \aap, 548, A10

\bibitem[\protect\citeauthoryear{Mosser et 
al.}{2012b}]{mos12b} Mosser, B., Goupil, M.~J., Belkacem, K., et al. 2012b, A\&A, 540, A143 

\bibitem[Mosser et 
al.(2014)]{mos14} Mosser, B., Benomar, O., Belkacem, K., et al.\ 2014, A\&A, 572, L5 

\bibitem[Mosser et al.(2015)]{mos15} Mosser, B., Vrard, M., Belkacem, K., et al.\ 2015, \aap, 584, A50

\bibitem[Mosser et al.(2017)]{mos17} Mosser, B., Pin{\c c}on, C., Belkacem, K., Takata, M., \& Vrard, M.\ 2017, \aap, 600, A1

\bibitem[Mosser et al.(2018)]{mos18} Mosser, B., Gehan, C., Belkacem, K., et al.\ 2018, \aap, 618, A109

\bibitem[Osaki(1975)]{osa75} Osaki, J.\ 1975, PASJ, 27, 237 

\bibitem[Pin{\c{c}}on et al.(2019)]{pin19} Pin{\c{c}}on, C., Takata, M., \& Mosser, B.\ 2019, \aap, 626, A125

\bibitem[Pin{\c{c}}on et al.(2020)]{pin20} Pin{\c{c}}on, C., Goupil, M.~J., \& Belkacem, K.\ 2020, \aap, 634, A68

\bibitem[Shibahashi(1979)]{shiba79} Shibahashi, H.\ 1979, PASJ, 31, 87 

\bibitem[Skilling(2004)]{sk04} Skilling, J.\ 2004, American Institute of Physics Conference Series, 735, 395 

\bibitem[Rogers et al.(1996)]{rog96} Rogers, F.~J., Swenson, F.~J., \& Iglesias, C.~A.\ 1996, ApJ, 456, 902

\bibitem[Takata(2005)]{tak05} Takata, M.\ 2005, PASJ, 57, 375

\bibitem[Takata(2006)]{tak06} Takata, M.\ 2006, PASJ, 58, 
893

\bibitem[Takata(2016a)]{tak16a} Takata, M.\ 2016a, PASJ, 68, 91

\bibitem[Takata(2016b)]{tak16b} Takata, M.\ 2016b, PASJ, 68, 109

\bibitem[\protect\citeauthoryear{Tassoul}{1980}]{tas80} Tassoul, M. 1980, ApJS, 43, 
469 

\bibitem[\protect\citeauthoryear{Unno et al.}{1989}]{unn89} Unno, W., Osaki, Y., Ando, 
H., Saio, H., 
\& Shibahashi, H. 1989, Nonradial oscillations of stars, Tokyo: University of Tokyo Press, 1989, 2nd ed.,

\bibitem[Vrard et al.(2016)]{vra16} Vrard, M., Mosser, B., \& Samadi, R.\ 2016, \aap, 588, A87

\bibitem[White et al.(2011)]{whi11} White, T.~R., Bedding, 
T.~R., Stello, D., et al.\ 2011, ApJ, 743, 161

\bibitem[\protect\citeauthoryear{White et al.}{2012}]{whi12} 
White T.~R., et al., 2012, ApJ, 751, L36




\end{thebibliography}
\end{document}